\documentclass[a4paper,fleqn,usenatbib]{mnras}

\usepackage[T1]{fontenc}
\usepackage{ae,aecompl}

\usepackage{graphicx}	
\usepackage{amsmath}	
\usepackage{amssymb}	
\usepackage{times}
\usepackage{xspace}

\usepackage{color}
\usepackage{hyperref}
\usepackage{supertabular}
\hypersetup{colorlinks, linkcolor=blue}

\def\eagle{\textsc{Eagle}\xspace}
\def\ceagle{\textsc{c-eagle}\xspace}
\def\subfind{\textsc{Subfind}\xspace}
\def\msun{{\rm M}_{\sun}}

\title[Building the MACS\,J0717 super-cluster]{
Growing a `Cosmic Beast': Observations and Simulations of MACS\,J0717.5+3745}
\author[M.\ Jauzac et al.\ 2018]
{M.\ Jauzac,$^{1,2,3,4}$\thanks{E-mail: mathilde.jauzac@durham.ac.uk}
D.\ Eckert,$^{5}$
M.\ Schaller,$^{2}$
J.\ Schwinn,$^{6}$
R.\ Massey,$^{1,2}$
Y.\ Bah\'e,$^{7,8}$
C.\ Baugh,$^{2}$
\newauthor D.\ Barnes,$^{9,10}$
C.\ Dalla Vecchia,$^{11,12}$
H.\ Ebeling,$^{13}$
D.\ Harvey,$^4$
E.\ Jullo,$^{14}$
S. T.\ Kay,$^{9}$
\newauthor J.-P.\ Kneib,$^{4,14}$
M.\ Limousin,$^{14}$
E.\ Medezinski,$^{15}$
P.\ Natarajan,$^{16}$
M.\ Nonino,$^{17}$ 
A.\ Robertson,$^{2}$
\newauthor S. I.\ Tam$^{1}$\& 
K.\ Umetsu$^{18}$ 
\\
\\
\\
$^{1}$Centre for Extragalactic Astronomy, Department of Physics, Durham University, Durham DH1 3LE, U.K.\\
$^{2}$Institute for Computational Cosmology, Durham University, South Road, Durham DH1 3LE, U.K.\\
$^{3}$Astrophysics and Cosmology Research Unit, School of Mathematical Sciences, University of KwaZulu-Natal, Durban 4041, South Africa\\
$^{4}$Laboratoire d'Astrophysique, Ecole Polytechnique F\'ed\'erale de Lausanne (EPFL), Observatoire de Sauverny, CH-1290 Versoix, Switzerland\\
$^{5}$Max-Planck Institut f\"{u}r Extraterrestrische Physik, Giessenbachstrasse 1, 85748 Garching, Germany\\
$^{6}$Zentrum f\"ur Astronomie, Institut f\"{u}r Theoretische Astrophysik, Universit\"{a}t Heidelberg, Philosophenweg 12, D-69120 Heidelberg, Germany\\
$^{7}$Leiden Observatory, Leiden University, PO Box 9513, 2300 RA Leiden, The Netherlands\\
$^{8}$Max-Planck-Institut f\"{u}r Astrophysik, Karl-Schwarzschild Str. 1, 85748 Garching, Germany\\
$^{9}$Jodrell Bank Centre for Astrophysics, School of Physics and Astronomy, The University of Manchester, Manchester M13 9PL, UK\\
$^{10}$Department of Physics, Kavli Institute for Astrophysics and Space Research, Massachusetts Institute of Technology, Cambridge, MA 02139, USA\\
$^{11}$Instituto de Astrof\'isica de Canarias, C/V\'ia L\'actea s/n, E-38205 La Laguna, Tenerife, Spain\\
$^{12}$Departamento de Astrof\'isica, Universidad de La Laguna, Av. del Astrof\'isico Francisco S\'anchez s/n, E-38206 La Laguna, Tenerife, Spain\\
$^{13}$Institute for Astronomy, University of Hawaii, 2680 Woodlawn Drive, Honolulu, Hawaii 96822, USA\\
$^{14}$Aix Marseille Univ, CNRS, LAM, Laboratoire d'Astrophysique de Marseille, Marseille, France\\
$^{15}$Department of Astrophysical Sciences, 4 Ivy Lane, Princeton, NJ 08544, USA\\
$^{16}$Department of Astronomy, Yale University, 260 Whitney Avenue, New Haven, CT 06511, USA\\
$^{17}$INAF- Osservatorio Astronomico di Trieste , Via Tiepolo 11, I-34131 Trieste, Italy\\
$^{18}$Institute of Astronomy and Astrophysics, Academia Sinica, P.O. Box 23-141, Taipei 10617, Taiwan
}

\date{Accepted XXX. Received YYY; in original form ZZZ}

\pubyear{2017}

\begin{document}
\label{firstpage}
\pagerange{\pageref{firstpage}--\pageref{lastpage}}
\maketitle

\begin{abstract}
We present a gravitational lensing and X-ray analysis of a massive galaxy cluster and its surroundings. The core of MACS\,J0717.5+3745 ($M(R<1\,{\rm Mpc})\sim$\,$2$$\times$$10^{15}\,\msun$, $z$=$0.54$) is already known to contain four merging components.
We show that this is surrounded by at least seven additional substructures with masses ranging from $3.8-6.5\times10^{13}\,\msun$, at projected radii $1.6$ to $4.9$\,Mpc. 
We compare MACS\,J0717 to mock lensing and X-ray observations of similarly rich clusters  in 
cosmological simulations.
The low gas fraction of substructures predicted by simulations turns out to match our observed values of $1$--$4\%$.
Comparing our data to three similar simulated halos, we infer a typical growth rate and substructure infall velocity. That suggests MACS\,J0717 could evolve into a system
similar to, but more massive than, Abell\,2744 by $z=0.31$, and 
into a $\sim$\,$10^{16}\,\msun$ supercluster by $z=0$.
The radial distribution of infalling substructure suggests that merger events are strongly episodic; however
we find that the smooth accretion of surrounding material remains the main source of mass growth even for such massive clusters.
\end{abstract}



\section{Introduction}
\label{sec:intro}
Massive galaxy clusters are the most massive gravitationally bound structures in the present Universe, having grown by repeatedly accreting smaller clusters and groups \citep[e.g.][]{fakhouri08,genel10}. 
However, most of the mass in the Universe is located outside gravitationally-bound halos. 
Clusters reside at the vertices of a cosmic network of large-scale filaments \citep{bond96}.
Numerical simulations predict these filaments contain as much as half of the Universe's baryons \citep{CO99,dave01} in the form of a warm plasma \citep{fang02,fang07,kaastra06,rasmussen07,galeazzi09,williams10,eckert15}, and the majority of of the Universe's dark matter \citep{aragoncalvo10}. 

Filaments are the scaffolding inside which clusters are built. They control the evolution of clusters.
Particularly in the outskirts of a galaxy cluster, filaments create preferred directions for the accretion of smaller halos, affecting the growth and shape of the main halo.
Filaments also channel infalling galaxies, accelerating or `pre-processing' their morphological and stellar evolution.
Substructures in filaments bias cluster mass measurements, especially from weak gravitational lensing \citep{martinet16}. 
Mis-calibrating cluster number counts can bias cosmological constraints \citep{martinet16}, and mis-calibrating clusters' magnification of background galaxies can bias high-redshift galaxy number counts by up to 30\% \citep{acebron17}. 
For all these reasons, observationally assessing substructures is essential if cluster evolution is to be understood and exploited.

One of the most efficient ways of mapping a distribution of mass dominated by dark matter is gravitational lensing: the bending of light from a background source as it passes near a foreground mass \citep[for reviews see ][]{massey10,KN11,hoekstra13}. Gravitational lensing is a purely geometrical effect, and is thus insensitive to the dynamical state of the cluster. It has been used extensively to probe the matter distribution in and around galaxy clusters \citep[e.g. ][]{kneib03,clowe04b,clowe06,bradac06,limousin07b,richard10,zitrin11,harvey15,massey15,jauzac16b,natarajan17,chirivi17}. Additionally, observations of X-ray emission from infalling structures reveal the presence of hot gas -- which, if virialized, is also an unambiguous signature of an underlying dark-matter halo \citep{neumann01,randall08,eckert14,eckert17,degrandi16,ichinohe15}. The combination of lensing and X-ray information is thus a powerful tool to study the processes governing the growth of massive galaxy clusters.

The {\em Hubble Space Telescope} (HST) has recently obtained the deepest ever images of galaxy clusters, through the \emph{Hubble Frontier Fields} programme \citep[HFF; ][]{lotz17}. 
This targets six of the most massive clusters in the observable Universe, which we call `cosmic beasts' because of their impressive size ($M_{200}\sim10^{15}~\msun$).
These objects are rare but, as extrema, are also ideal tests of the cosmological paradigm. 

One HFF galaxy cluster, Abell\,2744, has been the source of recent debate. 
At redshift $z=0.31$, it has a complex distribution of substructure in its core, and three filaments containing both dark matter and gas \citep{eckert15}. 
\cite{jauzac16b} recorded a total of seven $>$$5\times10^{13}\,\msun$ mass substructures, projected within $1.2$\,Mpc of the cluster centre. 
Searching in the Illustris simulation volume \citep{vogelsberger14}, \cite{natarajan17} could not find a mass analog to Abell\,2744. However, performing zoom-in simulations they generated a comparable mass cluster and found good agreement between the lensing derived subhalo mass function determined from the \cite{jauzac15b} strong-lensing mass reconstruction and that derived from the simulated cluster across three decades in mass from $10^{9} - 10^{12.5}\,\msun$.
\cite{schwinn17} were unable to find any systems as rich in substructures in the entire Millenium-XXL simulation \citep{angulo12}. 
However, they suggested numerical and observational caveats to explain this apparent inconsistency: reduced resolution of the \textsc{subfind} subhalo finder algorithm \citep{springel01,dolag09} at lower density contrasts in the core of the main halo, comparison between 3D \textsc{subfind} masses from simulations and 2D projected masses from lensing data, and the contamination of lensing masses by line-of-sight substructures.
\cite{mao17} argued that as lensing measurements integrate mass along a line of sight, they include mass from additional structures, and quantified this effect using the Phoenix cluster simulations \citep{gao12}.
The discrepancy might therefore be reduced by simulating observable quantities \citep{schwinn18}, or by simultaneously fitting  all the components of a parametric mass model. 

To obtain another example of the assembly of substructures, here we study an even more massive HFF galaxy cluster, MACS\,J0717.5+3745 (MACS\,J0717), at higher redshift, $z=0.54$.
This is the most massive galaxy cluster known at $z>0.5$ \citep{edge03,ebeling04,ebeling07}, and one of the strongest gravitational lenses known \citep{diego15a,limousin16,kawamata16}.
Lensing and X-ray analyses of the cluster core have revealed a complex merging system involving four cluster-scale components \citep{ma09,zitrin09b,limousin12}.
A single filament extending South-East of the cluster core has been detected in the 3D distribution of galaxies \citep{ma08} and the projected total mass from weak lensing \citep{jauzac12,medezinski13,martinet16}.
We now exploit recent, deep observations from the \emph{Hubble Space Telescope}, \emph{Chandra X-ray Observatory}, \emph{XMM-Newton X-ray Observatory}, Subaru and Canada-France-Hawaii telescopes, to map the distribution of substructure up to $\sim5$ Mpc from the cluster core in all directions, and to investigate the way the filament funnels matter into the centre.
We then compare our results to theoretical predictions from the MXXL and Hydrangea/C-EAGLE \citep{bahe17,barnes17} simulations.

This paper is organised as follows.
Section~\ref{sec:obs} presents the multi-wavelength datasets used in our analysis. 
Section~\ref{sec:simu} presents the numerical simulations used in our comparison.
Section~\ref{sec:lmodel} describes our gravitational lensing measurements, and Section~\ref{sec:allmodel} summarises the technique we use to combine strong- and weak-lensing information.
Section~\ref{sec:results} compares our lensing results to the distribution of X-ray emitting gas. 
Section~\ref{sec:compth} discusses our findings in the context of theoretical predictions from numerical simulations. 
We conclude in Section~\ref{sec:conclusion}.
For geometric calculations, we assume a $\Lambda$ cold dark matter ($\Lambda$CDM) cosmological model, with $\Omega_{m}=0.3$, $\Omega_{\Lambda}=0.7$, and Hubble constant $H_{0}=70$\,km\,s$^{-1}$\,Mpc$^{-1}$. 
Thus 1\, Mpc at $z=0.54$ subtends an angle on the sky of 2.62\arcmin, and at $z=0.31$ subtends 3.66\arcmin.
We quote all magnitudes in the AB system.

\section{Observations}
\label{sec:obs}

\subsection{Hubble Space Telescope (HST) imaging}
\label{sec:hst} 
The core of MACS\,J0717 was initially imaged by HST as part of the X-ray selected \emph{MAssive Cluster Survey} \citep[MACS; ][]{ebeling01}. 
Observations of 4.5\,ks were obtained in each of F555W and F814W passbands of the \emph{Advanced Camera for Surveys} (ACS) (GO-09722 and GO-11560; PI: Ebeling).
It was subsequently re-observed as part of the \emph{Cluster Lensing And Supernovae with Hubble} survey \citep[CLASH, GO-12066, PI: Postman; ][]{postman12}, for an additional 20 orbits across 16 passbands from the UV to the near-infrared, with ACS and the \emph{Wide-Field Camera 3} (WFC3).
Finally, the strong lensing power of MACS\,J0717 made it an ideal target for the \emph{Hubble Frontier Fields} observing campaign \citep[HFF, ][]{lotz17}. 
Its core was thus observed again for 140 orbits during Cycle 23, in 7 UV to near-infrared passbands, with ACS and WFC3 (GO-13498, PI: Lotz).

Meanwhile, a large-scale filament extending from the cluster core was discovered in photometric redshifts of surrounding galaxies from multi-colour ground-based observations \citep[][]{ebeling04,ma09}. 
This motivated mosaicked HST/ACS imaging of a $\sim$10$\times$20 arcmin$^{2}$ region around the cluster in F606W and F814W passbands during 2005 (GO-10420, PI: Ebeling).

Data reduction of the core images used the standard \textsc{hstcal} procedures with the most recent calibration files \citep{lotz17}. \textsc{astrodrizzle} was used to co-add individual frames after selecting a common ACS reference image using \textsc{tweakreg}. The final stacked images have a pixel size of 0.03\arcsec.
Data reduction of the mosaic observations is described in \cite{jauzac12}. This followed a similar procedure as the core observations, except that exposures were treated independently to avoid resampling of the images that could affect weak-lensing shape measurements. These final images also have a pixel size of 0.03\arcsec.

\subsection{Ground-based imaging}
Wide-field imaging around MACS\,J0717 has been obtained from the 8.2\,m Subaru telescope's SuprimeCam camera \citep[$34\arcmin\times27\arcmin$ field of view;][]{miyazaki02} in \emph{B}, \emph{V}, \emph{R$_{c}$}, \emph{I$_{c}$} and \emph{z'} bands \citep{medezinski13}. 
The 3.6\,m Canada-France-Hawaii Telescope (CFHT) has also obtained MegaPrime $u^{\ast}$-band imaging (1\,deg$^2$ field of view) and WIRCam $J$ and $K_{S}$-band  imaging.
All these data were reduced and analysed using standard techniques. For details, exposure times, and seeing conditions, we refer the reader to Table~2 in \cite{jauzac12} and Table~2 in \cite{medezinski13}.

These ground-based observations were used for two purposes: (1) to measure photometric redshifts to remove contamination from both foreground and cluster galaxies to the weak-lensing catalogues; and (2) to measure the shapes of background galaxies outside the region observed by HST, for the wide-field weak-lensing analysis.
Subaru weak lensing measurements were obtained from the \emph{R$_{c}$}-band image (see Sect.~\ref{sec:subwl} for more details).

\subsection{Chandra X-ray imaging}
The \emph{Chandra X-ray Observatory} has observed MACS\,J0717 on four occasions (OBSID 1655, 4200, 16235, and 16305), for a total exposure time of 243\,ks. All observations were performed in ACIS-I mode. We reduced the data using \textsc{ciao} v4.8 and \textsc{caldb} v4.7.2. We used the \texttt{chandra\_repro} pipeline to reprocess the event files with the appropriate calibration files and extracted source images in the [0.5-1.2] keV band using \texttt{fluximage}. We used the \texttt{blanksky} and \texttt{blanksky\_image} tools to extract blank-sky datasets to model the local background, and we renormalized the blank-sky data such that the count rate in the [9.5-12] keV band matches the observed count rate to take the long-term variability of the particle background into account \citep{HM06}. 

\subsection{XMM-Newton X-ray imaging}
\emph{XMM-Newton} has observed MACS\,J0717 three times (OBSID 067240101, 067240201, 067240301, PI: Million) for a total exposure time of 194\,ks. We reduced the data using \textsc{xmmsas} v15.0 and the corresponding calibration database. We used the Extended Source Analysis Software (\textsc{esas}) package \citep{snowden08} to analyze the data. We filtered the data for soft proton flares using the \texttt{pn-filter} and \texttt{mos-filter} tools, leading to a clean exposure time of 155\,ks for MOS and 136\,ks for pn. We extracted photon images in the [0.5-1.2] keV band for the three observations separately and used filter-wheel-closed data files to estimate the particle background contribution. Exposure times were computed using the \textsc{xmmsas} tool \texttt{eexpmap}, taking the vignetting curve of the telescope and CCD gaps into account. The images of the three EPIC instruments were then combined and the various observations were mosaicked to create a total image of the cluster and its surroundings.

We also extracted spectra of several regions (see Sect.~\ref{sec:lensingxstr}) to measure the thermodynamic properties of the gas. The spectra were extracted using the \textsc{esas} tasks \texttt{mos-spectra} and \texttt{pn-spectra}. Contaminating point sources were detected and excised using the \texttt{cheese} tool. Each background component was modeled separately and added to the total source model following the procedure described in \citet{eckert14}. The background is split between the non-X-ray background, which we model using a phenomenological model tuned to describe the spectral shape of the filter-wheel-closed data, and the sky background. The latter can be described as the sum of three components: \emph{(i)} an absorbed power law with a photon index of 1.46 to model the contribution of unresolved point-like sources \citep{deluca04}; \emph{(ii)} an absorbed thin plasma model with a temperature of 0.22\,keV to describe the X-ray emission of the Galactic halo \citep{mccammon02}; \emph{(iii)} a thin plasma model with a temperature of 0.11\,keV to model the local hot bubble. We used a source-free region located $\sim10$\,arcmin North-West of the cluster core to estimate the relative intensity of the three sky background components. The measured normalizations were then rescaled to the area of the regions of interest. Finally, the source spectra were modeled as a single-temperature APEC model \citep{apec}, leaving the temperature, emission measure and metal abundance as free parameters during the fitting procedure. For more details on the spectral modeling approach, we refer the reader to \citet{eckert14}.

\subsection{Spectroscopy \& Photometry}
MACS\,J0717 has been extensively surveyed with the Deep Imaging Multi-Object Spectrograph (DEIMOS), the Low Resolution Imaging Spectrometer (LRIS) and Gemini Multi-Object Spectrograph (GMOS), on the Keck-II, and Keck-I and Gemini-North telescopes respectively on Mauna Kea. 
These observations (detailed in \citealt{ma08} and summarised in \citealt{jauzac12}), cover both the core and the known filamentary structure.
The DEIMOS instrument set-up combined the 600ZD grating with the GC455 order-blocking filter, with a central wavelength between 6300 and 7000\,$\AA$. A total of 18 multi-object masks were observed with DEIMOS, with each of them having an exposure time of $\sim$3$\times$1800\,s, as well as 65\,s and 48\,s with LRIS and GMOS respectively. 
These spectroscopic observations yielded redshifts of 1079 galaxies, 537 of which are confirmed as cluster members.

\cite{ma08} presented a photometric redshift catalogue for galaxies with $m_{R_{C}} < 24.0$, compiled using the adaptive SED-fitting code \textsc{Le Phare} \citep[][]{arnouts99,ilbert06,ilbert09}. We use this to calibrate colour-colour selections and to estimate the contamination from foreground and cluster galaxies in the weak lensing catalogues.

\section{Numerical Simulations}
\label{sec:simu}
We use two state-of-the art cosmological simulations to establish theoretical expectations and to interpret our observational results. 

\subsection{The Millenium-XXL dark matter simulation}
\label{sec:mxxl}
The dark matter only Millenium-XXL simulation \citep[MXXL; ][]{angulo12} simulates the evolution of
dark matter in a $\Lambda$CDM Universe ($H_0~=~100h~=~73\,{\rm
km\,s^{-1}Mpc^{-1}}$, $\Omega_\Lambda = 0.75$, $\Omega_{\rm m}
= \Omega_{\rm dm} + \Omega_{\rm b} = 0.25$, $\Omega_{\rm b} = 0.045$
and $\sigma_8 = 0.9$). The dark matter fluid is traced by particles of
mass $m_{p} = 6.16 \times 10^{9}\, h^{-1}\msun$ within a cube of volume
($3\,h^{-1}{\rm Gpc}$)$^3$. Structures are detected within the MXXL
simulation on two hierarchical levels. Dark matter haloes are found
using the Friends-of-Friends algorithm \citep[FoF; ][]{Davis1985}
using a linking length of $b=0.2$. Within these FoF haloes,
gravitationally bound subhalos are identified using
the \textsc{subfind} algorithm \citep[][]{springel01,dolag09}.

\cite{schwinn17} searched the MXXL for an analogue of galaxy cluster Abell\,2744 ($z=0.31$), which contains 7 massive substructures at the cluster redshift plus one behind the cluster, within the central 2\ Mpc \citep{jauzac16b}. 
They found clusters as massive as Abell\,2744, but none with as many substructures -- at least not substructures detected by the \textsc{FoF} and \textsc{subfind} algorithms.
On the other hand, \cite{natarajan17} found good agreement with substructure in an Illustris zoom-in run with the strong-lensing derived substructure from the reconstruction of \cite{jauzac15b} between $10^{9}-10^{12.5}\,\msun$. However, they were unable to match the radial distribution of the observed substructures and they reported an excess at the massive end that they attributed to systematics arising from the choice of \textsc{subfind} as the halo finder.

However, further investigation using the particle data
of the MXXL, showed that this result seems to be caused by
different definitions of a subhalo in the \textsc{subfind} algorithm in
comparison to the gravitational lensing analysis \citep{schwinn18}. 
Due to the immense amount
of storage space needed, full MXXL particle data have only been stored for snapshots at $z=3.06$, $0.99$, $0.24$ and~$0$. Here, our comparison of MACS\,J0717 relies on the closest MXXL snapshot at $z = 0.24$. As we will show, by analyzing the particle data directly, we find two clusters with similar mass and a similar number of substructures (see Sect.~\ref{sec:simcomp} for details).

\subsection{The Hydrangea/C-EAGLE hydrodynamical simulation}
\label{sec:ceagle}
The Hydrangea/C-EAGLE suite of cluster simulations \citep{bahe17,barnes17} is a factor of a 1000 better in mass resolution than MXXL, and includes baryonic physics self-consistently. These 30 zoom-in simulations used the same physical model, resolution
and cosmology as the \eagle simulations \citep{schaye15,crain15},
making this the largest sample of high-resolution clusters currently
available. The clusters were selected from a parent dark matter only
simulation of side-length $3.2~{\rm Gpc}$ \citep{barnes16} using the
$\Lambda$CDM cosmological parameters derived from the 2013 analysis of
the \emph{Planck} data \citep{planck13_cosmo}
($H_0~=~100h~=~67.77\,{\rm km\,s^{-1}Mpc^{-1}}$, $\Omega_\Lambda =
0.693$, $\Omega_{\rm m} = 0.307$, $\Omega_{\rm b} = 0.04825$,
$\sigma_8 = 0.8288$, $n_s = 0.9611$ and $Y=0.248$). As in the MXXL
case, gravitationally bound halos were found in the simulation using
the FoF and {\sc subfind} algorithms.  At $z=0$ this simulation volume
contains $>180,000$ halos with $M_{200}>10^{14}~\msun$.
Haloes that were
not the most massive object within a radius of $30~{\rm Mpc}$ or
$20\,R_{200}$ (which ever is larger) around their centre were 
removed from the sample, and 30 were selected for zoom-in re-simulation \citep[see ][]{bahe17}.

Higher resolution zoom-in initial conditions for each halo were then
generated at $z=127$ based on second-order perturbation theory
following the method of \cite{jenkins10}. The initial particle masses
were set to $m_{\rm DM}=9.7\times10^6~\msun$ and $m_{\rm
g}=1.8\times10^6~\msun$ for the dark matter and gas
respectively. The Plummer-equivalent softening length was set to
$0.7~{\rm kpc}$ at $z<2.8$ and is fixed in comoving space to
$2.66~{\rm kpc}$ at higher redshift.

The initial conditions were then run using the \eagle simulation
code \citep{schaye15,crain15}. The code is a highly modified version
of the Tree-PM/SPH code {\sc Gadget} \citep{gadget_paper}. The
modifications to the hydrodynamics solver, including the use of the
Pressure-Entropy formulation of SPH \citep{hopkins13}, are described
by \cite{schaller15} and the subgrid physics modules were designed and
calibrated to reproduce the observed stellar mass function of galaxies
at low redshift, yield galaxy sizes in agreement with low-redshift
observations and a galaxy stellar mass - black hole mass relation
compatible with observed data \citep{crain15}. The galaxy formation
subgrid modules include metal-line cooling \citep{wiersma09a} from an
homogeneous \cite{haardt01} X-ray/UV background radiation
(with \ion{H}{}reionisation at $z = 11.5$), metallicity-dependent star
formation \citep{schaye04,schaye08}, metal
enrichment \citep{wiersma09b}, feedback from star
formation \citep{dallaVecchia12}, supermassive black-hole formation,
and AGN feedback \citep{booth09, rosasguevara15}. Post-processed
halo and sub-halo catalogues have then been generated for all output
redshifts using the \textsc{subfind} algorithm. The $z=0$ properties
of these 30 haloes are given in appendix A1 of \cite{bahe17} whilst
derived X-ray observable properties can be found in appendix A1
of \cite{barnes17}. All halos were also simulated at the same
resolution without baryonic processes.

\section{Gravitational lensing measurements}
\label{sec:lmodel}

\subsection{Strong-lensing constraints}
\label{sec:slmodel}
The deep HFF observations dramatically improved the strong-lensing mass model of the core of MACS\,J0717 \citep{zitrin09b,limousin12} thanks to the identification of more than 200 multiple images \citep[][]{diego15a,limousin16,kawamata16}.
For this analysis, we use \cite{limousin16}'s mass model, which includes 51 multiple image systems (a total of 132 multiple images) to constrain the mass distribution of the cluster, 10 of which are spectroscopically confirmed. Ideally one would like a spectroscopic redshift confirmation for each systems, but this is unfortunately not possible as we do not have unlimited access to telescope time. However with MACS\,J0717 we are able to sample the redshift space behind the cluster thanks to the 10 systems with spectroscopic confirmation, thus decreasing the impact of the mass sheet degeneracy to the \cite{limousin16} model.
\cite{johnson16} investigated the impact of the lack/abundance of spectroscopic redshifts on the resulting accuracy of the mass
model and showed that at least a few multiple image systems with spectroscopic confirmations are crucial to produce a reasonable estimate of the mass (and the magnification). They also show that the availability of numerous (>15) spectroscopically confirmed multiple image systems increases the accuracy of the lens model without necessarily further improving the precision of the mass model. We refer the reader to the published work by \cite{johnson16} for further details.
The best-fit mass model comprises four cluster-scale halos, which are coincident with the four main light peaks, plus 90 galaxy-scale halos in order to account for the impact of cluster galaxies on the geometry of nearby multiple images \citep[][]{natarajan97}. 
These galaxy-scale halos correspond to cluster member galaxies identified with spectroscopic and photometric redshifts.

\cite{limousin16} presented two alternative strong-lensing mass models: one named \emph{cored}, which has a relatively flat distribution of mass in the smooth component, and one named \emph{non-cored} which results in a more ``peaky'' mass distribution. Both models reproduce the geometry of the multiple images almost equally well, with an RMS offset between observed and predicted positions of images of 1.9\arcsec and 2.4\arcsec for the \emph{cored} and \emph{non-cored} models respectively.
We tested both strong-lensing models in our strong+weak-lensing analysis. Both give similar results, as expected. However, for simplicity we shall only quote the combination of the \emph{cored} strong-lensing mass model with our weak-lensing constraints in this paper.

\subsection{HST weak-lensing catalogue}
\label{sec:hstwl}
We note that we do not use HFF data for the weak-lensing analysis as it only covers the core of MACS\,J0717 that is highly spatially extended and thus dominated by strong-lensing. Our HST weak-lensing analysis therefore relies on moderate depth HST/ACS imaging from the mosaic presented in Sect.~\ref{sec:hst}.
We measure the weak gravitational lensing shear signal from the shapes of galaxies in the ACS/F814W band. 
Our method is based on the HST/ACS lensing pipeline developed by \cite{leauthaud07} for COSMOS and adapted to galaxy clusters by \cite{jauzac12}. This shear catalogue has already been published in \cite{jauzac12}, so here we provide only a short summary of the procedures.
\subsubsection{Background galaxy selection}
We first detect sources using \textsc{sextractor} \citep{BA96}, employing the `hot-cold' method \citep{rix04,leauthaud07} optimised for the detection of faint objects. This catalogue is then cleaned to remove spurious and duplicate detections. The star-galaxy classification is performed by looking at the distribution of sources in the magnitude (\textsc{mag$\_$auto}) versus peak surface brightness (\textsc{mu$\_$max}) plane.

Only the images of galaxies behind the cluster have been gravitationally lensed by it.
Foreground galaxies and cluster members must be removed from the shear catalogue, otherwise they will dilute the measured shear signal.
For the 15\% of galaxies detected by HST that have spectroscopic or reliable photometric redshifts \citep{ebeling14}, separating these galaxy populations is easy.
For the remaining $\sim$85\% of galaxies, we apply a $(B-V)$ versus $(u-B)$ colour-colour selection, which is calibrated using the spectroscopic and photometric redshifts in the rest of the catalogue. 
Selection criteria for photometric redshifts, and a detailed discussion of colour-colour selections and their calibration is provided in Section~3.2 of \cite{jauzac12}.

\subsubsection{Galaxy shape measurements}
We used the RRG moment-based shear measurement method \citep{rhodes00} to measure the shape of \emph{HST}-detected background galaxies. 
This was specifically developed for space-based data with a small, diffraction-limited point-spread function (PSF). It reduces noise by linearly correcting each shape moment for the effect of PSF convolution, and only dividing moments to compute an ellipticity at the very end.
Both the size and the ellipticity of the ACS PSF vary considerably with time, due to `breathing' of the telescope. Thermal fluctuations as parts of the telescope pass in and out of sunlight continually adjust its effective focus, thus making the PSF larger and more circular.
To model the PSF we used the grid of simulated PSF at varying focus offset created by \cite{rhodes07} using \textsc{tinytim} 6.3.

RRG returns a measure of each galaxy's apparent size, $d$, and apparent ellipticity, represented by a vector $\textbf{e} = (e_{1}, e_{2})$. 
From the latter, we obtain a shear estimator, $\tilde{\gamma} = C {\bf e}/G$, where $G$ is the shear polarizability (which is computed from higher order shape moments of a large sample of galaxies), and $C=1/0.86$ is a calibration factor computed by running the algorithm on mock HST data containing a known signal \citep{leauthaud07}.

\subsubsection{Catalogue cuts and weighting}
We exclude from the catalogue any galaxies with shape parameters that our experience running the RRG algorithm on mock data suggests may be unduly noisy or biased. 
We keep only galaxies with detection significance $S/N > 4.5$; ellipticity $|\textbf{e}|< 1$; and size $d>0.13\arcsec$.
Although ellipticity is by definition lower or equal to 1, RRG allows measured values greater than 1 because of noise. The restriction on the size of the galaxy is intended to eliminate sources with sizes approaching that of the PSF, thus making the shape of the galaxy difficult to measure.

Following \cite{leauthaud10}, we also use an inverse-weighting scheme to optimize overall signal-to-noise from the remaining galaxies.
We estimate the uncertainty in each shear estimator, $\sigma_{\tilde{\gamma}}$, by adding in quadrature intrinsic shape noise, $\sigma_{\mathrm{int}}$, plus shape measurement error, $\sigma_{\mathrm{meas}}$. 
We assume that intrinsic shape noise $\sigma_{\mathrm{int}}=0.27$, and errors on each ellipticity component are obtained by linearly propagating the covariance matrix of the moments \citep{leauthaud10}.  Weights $w_{\tilde{\gamma}} = 1/\sigma_{\tilde{\gamma}}^{2}$ then suitably down-weight the impact of noisy, faint galaxies.

In order to ensure unbiased measurements when combining strong- and weak-lensing information, we finally remove 
all galaxies located in the multiple-image region.
Our final HST weak-lensing catalogue consists of 10\,170 background galaxies, corresponding to a density of $\sim$52\,galaxies per arcmin$^{2}$.

\subsection{Subaru weak-lensing catalogue}
\label{sec:subwl}
In survey regions not covered by HST imaging, we measure the weak gravitational lensing shear signal from the shapes of galaxies in Subaru \emph{R$_{c}$}-band imaging. 
Our shear catalogue has already been published in \cite{medezinski13}, so here we provide only a short summary of the procedures.
\subsubsection{Galaxy shape measurements}
Our wide-field weak-lensing analysis uses the shape catalog obtained
by the CLASH collaboration \citep[][]{postman12} from deep
multi-band Subaru/Suprime-Cam ($BVR_\mathrm{c}i'z'$) and CFHT
(MegaPrime $u^*$ and WIRCam $JK_\mathrm{S}$) observations. Full
details of the image reduction, photometry, weak-lensing shape
analysis, and background source selection are given in \cite{medezinski13} and \cite{umetsu14} \citep[see their Section 4; for more
details on weak-lensing systematics, see Section 3 of ][]{umetsu16}. Briefly summarizing, the weak-lensing analysis procedures
include (1) object detection using the \textsc{imcat} peak finder \citep[][]{KSB95}, {\sc hfindpeaks}, (2) careful close-pair rejection to
reduce the crowding and deblending effects, and (3) shear calibration
developed by \cite{umetsu10}. For each galaxy a
shear calibration factor of $1/0.95$ is included to account for the residual
correction estimated using simulated Subaru/Suprime-Cam images \citep[][]{umetsu10}. The CLASH Subaru shape measurements used the
Suprime-Cam $R_\mathrm{c}$ data, which have the best image quality
among the data in terms of the stability and coherence of the
PSF-anisotropy pattern, and were taken in fairly good seeing
conditions \citep[0.79\arcsec in $R_\mathrm{c}$; see Table~2 of ][]{medezinski13}.

\subsubsection{Background galaxy selection}
Following \cite{medezinski10}, we identify background galaxies using a colour-colour selection in the ($B - R_{c})$ versus ($R_{c}-z'$) plane calibrated with evolutionary tracks of galaxies \citep[for more details see][]{medezinski10,umetsu10} and the COSMOS deep photometric-redshift catalogue \citep[][]{ilbert09}.
Three samples are identified in this colour-colour space: red, blue and green samples. The green sample encompasses mainly cluster members, and the red and blue ones two distinct lensed galaxy populations.
While the red sample is limited to a magnitude $mag_{z'}<25$, the blue sample extends to fainter magnitude, $mag_{z'}<26$, as the number density of bluer galaxies grows significantly higher with magnitude. We adopt conservative colour limits, in order to limit signal dilution due to the presence of cluster galaxies and foreground objects.

Our final Subaru weak-lensing catalogue consists of 4856 and 4738 galaxies in the red and blue lensed samples respectively. 
This correspond to a density of 9.6\,galaxies per arcmin$^{2}$ and 11.5\,galaxies per arcmin$^{2}$ throughout the SuprimeCam field of view.

\section{Mass Modelling}
\label{sec:allmodel}

\subsection{Strong+Weak lensing with Lenstool}
The combination of strong- and weak-lensing constraints follows the methodology described in \cite{jauzac15a} and \cite{jauzac16b}. We refer the reader to these publications for detailed discussions, and here only summarize the technique. 
It consists of combining both the parametric and the non-parametric approaches in the \textsc{lenstool} software \citep[][]{jullo07,jullo09,jauzac12,jullo14} in order to accommodate the high precision possible in the core thanks to strong-lensing constraints, while allowing more flexibility in the outskirts due to the lower information-density of the weak-lensing shear signal.
We thus keep the strong-lensing parametric model described in Sect.~\ref{sec:slmodel} fixed to its best-fit values, and add a multi-scale grid of radial basis functions (RBF) outside the cluster core to fit the weak-lensing constraints while optimizing the RBF's amplitudes. Such an approach allows us to appropriately weight the strong-lensing constraints \citep[see][]{jauzac15a}.

The parametric model is composed of four cluster-scale haloes \citep{limousin16} to which we add 2244 pseudo-isothermal elliptical mass distribution potentials \citep[PIEMD;][]{eliasdottir07} that represent the member galaxies, and a multi-scale grid of 2630 RBFs.
Each RBF is modeled by a truncated isothermal sphere with core potential. Its position is fixed, and only its amplitude is allowed to vary over the optimization process. Its core radius, $s$, is set to the distance to its closest neighbor, and its cut radius, $t$, is assumed to be $3\times s$ \citep[][]{jullo09}. 
The optimal solution we found consists of a multi-scale grid composed of 2630 RBFs, with $s=24$\arcsec for the smallest RBFs in regions with HST imaging (see Sect.~\ref{sec:gridres}): a maximum resolution similar to that obtained by \cite{jauzac12}.
Outside this field, where the density of background galaxies is the lowest due to the absence of high-resolution imaging from HST, the RBF's core radii vary between $s=192$\arcsec and $s=383$\arcsec. 
Computational limitations currently prevent \textsc{lenstool} from simultaneously optimising the grid and the physical properties of individual cluster galaxies.
The cut radius, ellipticity and velocity dispersion of the galaxy-scale halos are thus fixed to the values obtained by \cite{limousin16}, and scaled from their luminosity in the $K$-band \citep[see][for further details]{jauzac12}.
These choices are considered reasonable as \textsc{lenstool} was tested on simulated clusters and was found to perform successfully at constraining scaling-relation parameters for the overall cluster galaxy population \citep{meneghetti17}.
Our team is working on overcoming those computational limitations and hopes to soon provide the community with an algorithm capable of fully optimizing all scales and lensing regimes with a non-parametric approach.

The contribution of the components of our model can be described as follows:
\begin{equation}
{\bf \tilde{\gamma}} = M_{\gamma \nu} {\bf \nu} + {\bf {\gamma}_{param}} + {\bf \sigma_{\tilde{\gamma}}} \, .
\end{equation}
where the vector ${\bf \nu}$ contains the amplitudes of the 2630 RBFs, the vector ${\bf \tilde{\gamma}}$ is defined in Sect.~\ref{sec:hstwl} and contains the individual shape measurements of the background galaxies, and ${\bf \gamma_{param}}$ is the fixed ellipticity contribution from the strong-lensing parametric model. ${\bf \sigma_{\tilde{\gamma}}}$ represents the noise as defined in Sect.~\ref{sec:hstwl}. 
$M_{\gamma\nu}$ is the transformation matrix which contains the cross-contribution of each individual RBF to each individual weak-lensing galaxy. For the two shear components, we can write the elements of $M_{\gamma\nu}$ as:
\begin{eqnarray}
\label{eq:dshear1}
\Delta_{1}^{(j,i)} &= &\frac{D_{LSi}}{D_{OSi}}\ \Gamma_{1}^i(|| \theta_i - \theta_j ||,\ s_i,\ t_i) \ , \\
\Delta_{2}^{(j,i)} &= &\frac{D_{LSi}}{D_{OSi}}\ \Gamma_{2}^i(|| \theta_i - \theta_j ||,\ s_i,\ t_i) \ .
\end{eqnarray}
where $\Gamma_{1}$ and $\Gamma_{2}$ are given in \citet[][Equation\ A8]{eliasdottir07}.
Note that the shear in the cluster core can be large, and thus the assumption from equation~\ref{eq:dshear1} may not be strictly valid. However, the contribution to the grid-based model originates primarily from the weak-lensing regime as the cluster core contribution is accounted for mainly by the strong-lensing parametric model. 

The parameter space is sampled using the \textsc{MassInf} algorithm implemented in the Bayesys library \citep{skilling98} which is itself implemented in \textsc{lenstool} \citep[][]{jullo07,jullo14}. At each iteration the most significant RBFs are identified, and their amplitude is then adjusted to fit the ellipticity measurements. As an output, the algorithm gives us a large number of Monte Carlo Markov Chain (MCMC) samples from which we can then derive mean values and errors on several quantities such as the mass density field and the magnification field amongst others.

Concerning the redshift of the background population, we follow the approach of \cite{jauzac15a} and \cite{jauzac16b}. For background galaxies that do not have a spectroscopic redshift or a secure photometric redshift, we assume a redshift distribution described by 
$\mathcal{N}(z) \propto e^{-(z/z_{0})^{\beta}}$, 
with $\beta = 1.84$ and a median redshift $\langle z\rangle = 1.586$ \citep{natarajan97,gilmore09}.
\textsc{lenstool} requires each source to have its own redshift.
Thus the redshifts for all galaxies without spectroscopic or photometric redshifts are randomly drawn from this distribution during the initialization phase.

\subsection{Grid resolution}
\label{sec:gridres}
Before converging on a grid of 2630 RBFs, we tested several possibilities including higher and lower resolution multi-scale grids, as well as uniform grids. Our main goal is to study the distribution of substructure in the outskirts of MACS\,J0717, so we need to be careful to not introduce spurious substructures due to a high level of noise in the grid. 
A second point to consider is the different density of background galaxies resolved in the HST and Subaru weak-lensing catalogues.

A baseline for this study is provided by the analysis of \cite{jauzac12}, which used HST weak-lensing data only. They tested the grid parameters, and converged on an optimal solution consisting of a multi-scale grid with $s=26$\arcsec for the smallest RBFs. 
For the present work to recover the filamentary structure with a similar significance level, we tried a uniform grid with a resolution of 24\arcsec. The motivation behind the uniformity of the grid is to avoid any prior on the mass distribution, such as that light traces mass. A uniform grid recovers the filament and all the substructures presented in Sect.~\ref{sec:lensingstr}. However, spurious detections are obtained due to a higher level of noise in the Subaru region as the resolution of the grid is too high compared to the density of weakly-lensed galaxies. Therefore, we tested a multi-scale grid to account for the non-uniform background galaxy density. The optimal solution we found consists of a multi-scale grid of 2630 RBFs, with the smallest RBFs having a core radius of $s=24$\arcsec in the HST field of view, and with RBF's core radii between $s=192$\arcsec and $s=383$\arcsec in the Subaru field of view.
Our choice is conservative as we limit ourselves to the high-mass substructures, avoiding over-extrapolation of the data that might lead to incorrect results.

\begin{figure}
\hspace*{-5mm}\includegraphics[width=0.5\textwidth]{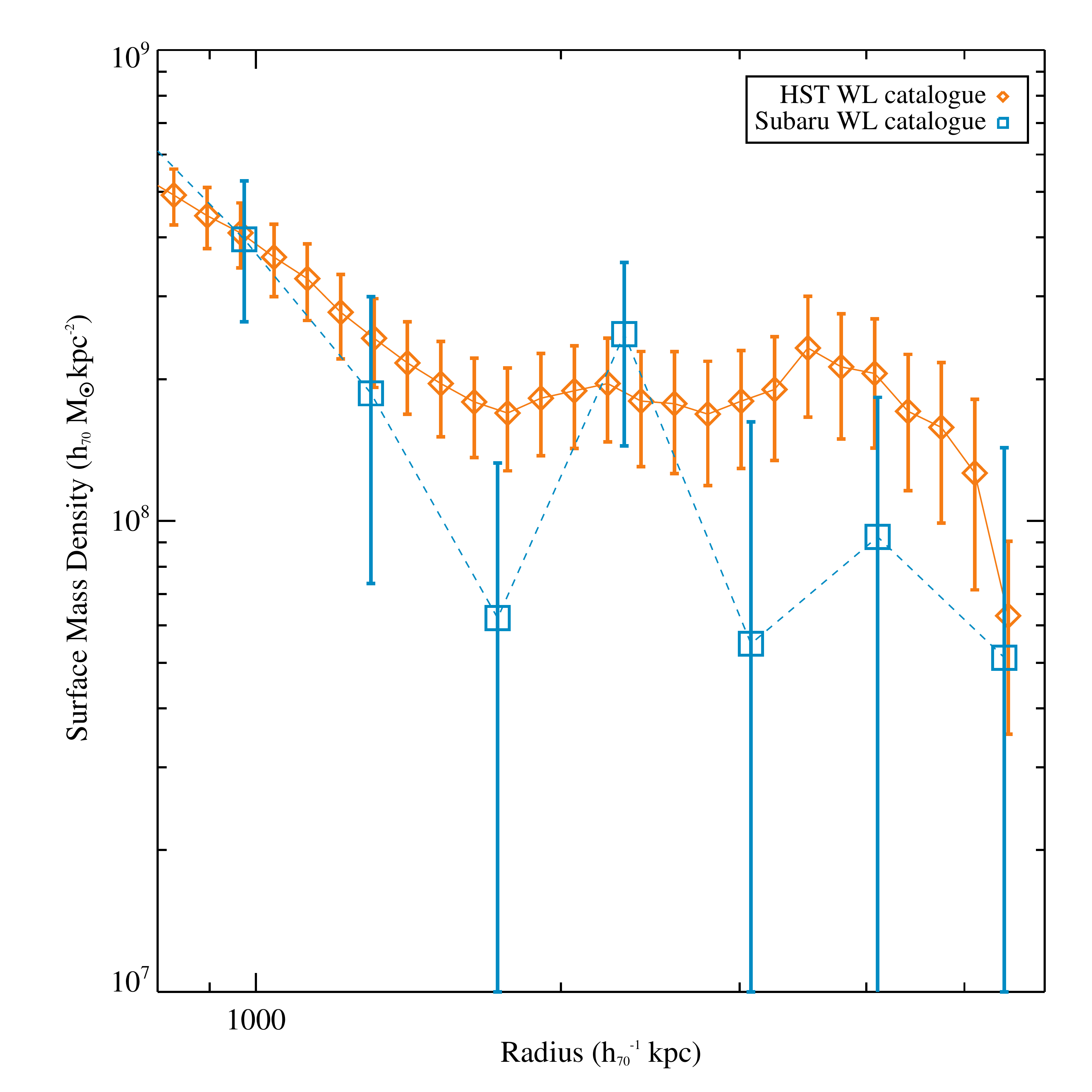}
\caption{
Density profiles from the Subaru (cyan squares) and HST (orange diamonds) weak-lensing analysis along the large-scale filament detected by \citet{jauzac12}. 
}
\label{subhst_prof}
\end{figure}
\section{Results}
\label{sec:results}

\subsection{Substructure detection from gravitational lensing}
\label{sec:lensingstr}
\begin{figure*}
\includegraphics[width=\textwidth]{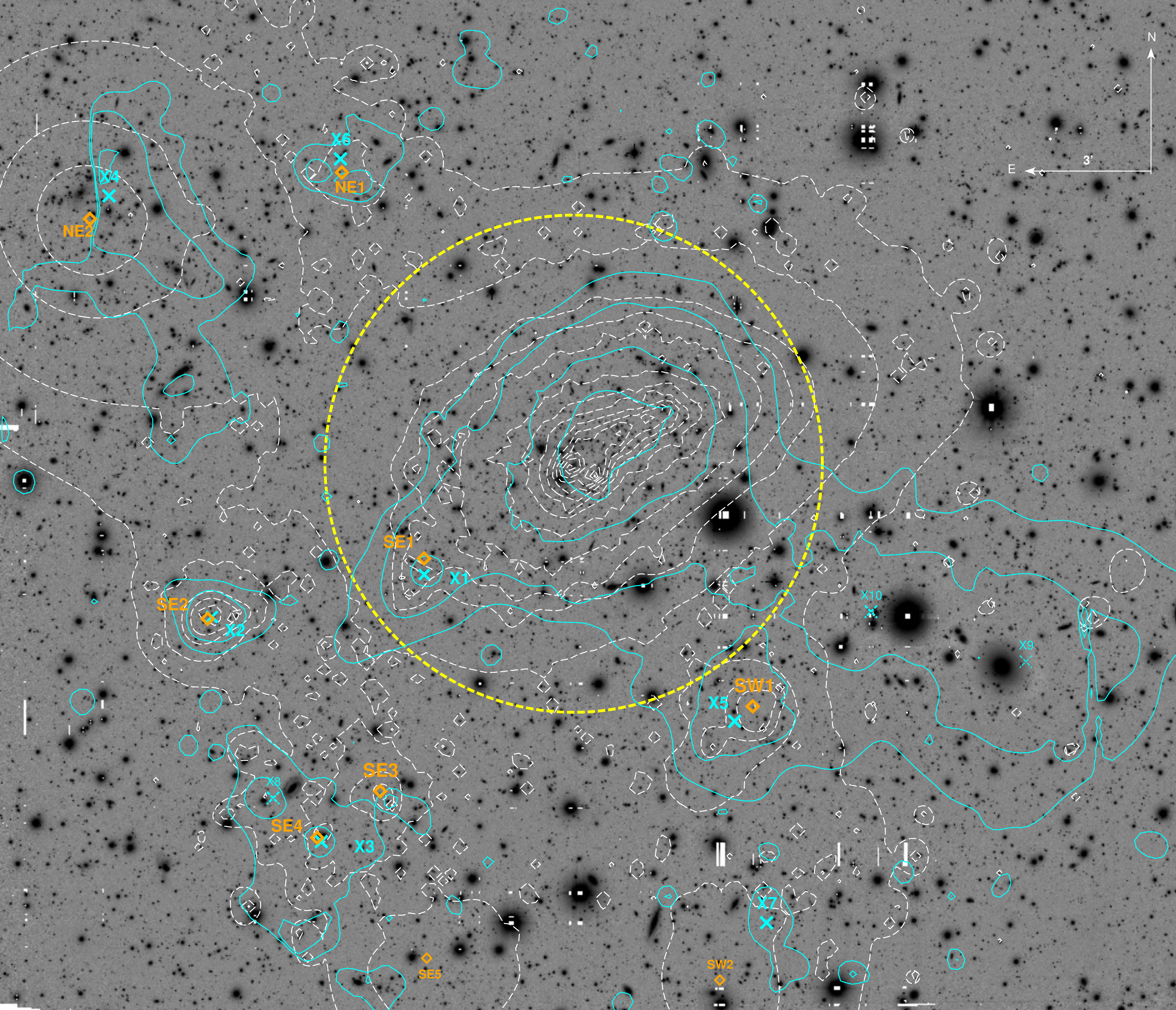}
\caption{
Subaru R-band image of MACS\,J0717. 
Orange diamonds highlight the position of substructures detected in the strong+weak lensing mass map (and listed in Table~\ref{tab_substr}); cyan crosses highlight the positions of remnant cores detected in the \emph{Chandra} and \emph{XMM-Newton} maps. White contours show the mass distribution derived from our strong+weak lensing mass model; 
cyan contours represent the gas distribution deduced from \emph{XMM-Newton} observations. The yellow circle has a radius of $R_{200}=2.3$\,Mpc (5.8').
}
\label{m0717_colour}
\end{figure*}
Our strong+weak lensing mass reconstruction of MACS\,J0717 reveals 9 substructures located between 1.6 and 4.9\,Mpc in projection from the cluster core ($\alpha$: 109.39820; $\delta$: 37.745778), which is itself composed of four merging clusters. Table~\ref{tab_substr} lists the substructures' and the cluster core (\emph{Core}) coordinates, masses, and detection significance.
All substructures are also highlighted in Fig.~\ref{m0717_colour} with orange diamonds. We note that the large-scale filament detected in \cite{jauzac12} is not illustrated clearly in Fig.~\ref{m0717_colour}. The structure is detected with 3$\sigma$ significance, a similar level as in \cite{jauzac12}. However, we chose to draw contours that highlight the substructure detections, rather than the lower-density filament. 
Moreover, as a test of consistency between the Subaru and the HST weak-lensing analysis we compare the density profiles obtained in this region as shown in Fig.~\ref{subhst_prof}. Both profiles show a good agreement, with the Subaru one having larger error bars due to a much lower density of background galaxies.

The \emph{Core} of MACS\,J0717 has been extensively studied due to its rich dynamical status, and therefore its lensing power. Its four main components are not the subject of this analysis, and are therefore all imbedded in the \emph{Core} component (see Sect.~\ref{sec:slmodel}). To test the reliability of our mass measurements, we first compare our mass values with published strong-lensing estimates from \cite{limousin16}, \cite{diego15a} and \cite{kawamata16}. For their \emph{core} model, \cite{limousin16} measure a total mass of $M_{L16}(R<990\,{\rm kpc}) = (2.229\pm 0.022)\times 10^{15}~\msun$, which is in excellent agreement with our measurement, $M(R<990\,{\rm kpc}) = (2.214\pm0.050)\times 10^{15}~\msun$.
\cite{diego15a} used a free-form method to build the strong-lensing mass model of MACS\,J0717 \citep[WSLAP+,][]{diego05,diego07,ponente11,lam14,sendra14}, and measured a mass of $M_{D15}(R<80\,{\rm kpc}) = 4.25\times 10^{13}~\msun$ (priv. comm.), which is in excellent agreement with our mass estimate within the same aperture, $M(R<80\,{\rm kpc}) = (4.24\pm0.03)\times 10^{13}~\msun$.
\cite{kawamata16} used the parametric \textsc{glafic} algorithm \citep{oguri10} and measured a total mass $M_{K16}(R<80\,{\rm kpc}) = 4.69\times 10^{13}~\msun$ (priv. comm.). While their estimate is slightly higher than ours, it is of the same order.

We now discuss the several substructures detected on our strong- and weak-lensing mass map. \emph{SE1} and \emph{SE2} were previously detected in \cite{jauzac12}, and as described in Sect.~\ref{sec:lensingxstr} both have an X-ray counterpart. They are also the most massive substructures found in MACS\,J0717 outskirts. 
The \emph{SE5}, \emph{NE2} and \emph{SW2} substructures are all at the edge of the mass map. It is therefore difficult to disentangle between real substructures and artifacts from the mass modeling technique. Nevertheless, if they are real, the mass estimates should be taken with care. \emph{NE2} is discussed in more detail in Sect.~\ref{sec:lensingxstr}.
Finally, \emph{SE1}, \emph{SE2}, \emph{SE3}, \emph{SE4} (and possibly \emph{SE5}) are all embedded in the large-scale filament identified in \cite{jauzac12}. Moreover, all the detected substructures show an optical counterpart, and appear to be at the redshift of the cluster when identifying their galaxy counterparts using photometric and spectroscopic redshifts from \cite{ma08}.
\emph{NE1}, \emph{NE2} and \emph{SE2} are all three detected by \cite{durret16}. This double identification confirms all three are at the cluster's redshift, as \cite{durret16} detected them as overdensities of red-sequence galaxies.

\begin{table*}
\caption{Coordinates, masses within 150\,kpc  and 250\,kpc apertures, significance of detection and projected distance to the cluster centre ($D_{C-S}$) for the substructures detected in the field of MACSJ\,0717.
We take the \emph{Core} coordinates as the one of the cluster itself following \citet{limousin16}, located close to the centre of Group C (see their Fig. 2). 
$^{\ast}$ These substructures are located at the edge of the grid, therefore their detection as well as their mass estimates should be taken with care.
}
\begin{center}
\begin{tabular}[h!]{lcccccc}
\hline
\hline
\noalign{\smallskip}
$ID$ & R.A.\ (deg)& Dec.\ (deg) & $M_{150}~[10^{13}~\msun]$ & $M_{250}~[10^{13}~\msun]$ & $\sigma$ & $D_{C-S}~[\rm{Mpc}]$ \\
\hline
\emph{Core} & 109.3982 & 37.745778 & $11.98 \pm 0.11$ & $30.78\pm 0.32$ & 130 & -- \\
\emph{SW1} & 109.3087625 & 37.6497725 & $ 2.41\pm 0.59$ & $ 6.19\pm 1.16$ & 5 & 2.8 \\
\emph{SW2}$^{\ast}$ & 109.3252847 & 37.54148293 & $1.34\pm$ 0.51& $3.84\pm 1.38$ & 3 & 4.9\\
\emph{SE1} & 109.4729667 & 37.70826611 & $2.28 \pm 0.24$ & $ 6.41\pm 0.62 $ &10 & 1.6 \\
\emph{SE2} & 109.58105 & 37.68432278 & $2.62 \pm 0.60$ & $ 6.51\pm 0.95$ & 8 & 3.6 \\
\emph{SE3} & 109.49475 & 37.61619444 & $2.20 \pm 0.55$ & $ 5.70\pm 1.31$ & 5 & 3.5 \\
\emph{SE4} & 109.5261625 & 37.59775361 & $ 1.85\pm 0.51$ & $4.46\pm 0.97$ & 4 & 4.1 \\
\emph{SE5}$^{\ast}$ & 109.4714417 & 37.5501475 & $1.74\pm 0.54$ & $4.72\pm1.48$ & 3 & 4.7\\
\emph{NE1} & 109.5142708 & 37.86093833 & $1.44 \pm0.46$ & $4.18 \pm 1.65$ & 3 & 3.4 \\
\emph{NE2}$^{\ast}$ & 109.6404125 & 37.84233667 & $ 2.27\pm 0.71$ & $6.44 \pm 2.00$ & 3 & 4.9 \\
\noalign{\smallskip}
\hline
\hline
\end{tabular}
\label{tab_substr}
\end{center}
\end{table*}

\begin{table*}
\caption{\label{tab:remnant}Coordinates, temperatures, X-ray luminosities in the [0.5-2] keV band, gas masses (within $\sim$250\,kpc) and lensing counterpart (if any) of the infalling structures identified in our X-ray analysis. $^{\ast}$ X8 corresponds to foreground spiral galaxy that we identified as 2MASSXJ\,07180932+3737031. $^{\ast\ast}$ X9 corresponds to a well-know submilimetre galaxy, 2MASSXJ\,07164427+3739556, at a redshift $z=0.06907$. $^{\ast\ast\ast}$ X10 location matches with a possible foreground galaxy, however we could not find any redshift.}
\begin{center}
\begin{tabular}{cccccccc}
\hline
\hline
ID & R.A. (deg) & Dec. (deg) & kT [keV] & $L_{X,250}$ [$10^{42}$ erg s$^{-1}$] & $M_{\rm gas,250} [10^{11}~\msun]$ & $S_{lensing}$ & $f_{gas,250}$
\\
\hline
X1 & 109.47288 & 37.701895 & $3.42\pm 0.18$ & $23.0\pm1.2$ & $24.6\pm1.0$ & \emph{SE1} & 0.04 \\
X2 & 109.57894 & 37.685011 & $1.82\pm 0.26$ & $10.7\pm0.9$ & $16.5\pm1.3$ & \emph{SE2} & 0.03 \\
X3 & 109.52414 & 37.596199 & $1.60\pm0.29$ & $11.0\pm0.9$ & $13.7\pm1.8$ & \emph{SE4} & 0.03\\
X4 & 109.63088 & 37.851494 & $1.52\pm0.16$ & $18.0\pm4.2$ & $16.6\pm3.5$ & \emph{NE2} & 0.03\\
X5 & 109.31781 & 37.643859 & $1.01\pm0.11$ & $5.6\pm0.8$ & $8.3\pm3.2$ & \emph{SW1} & 0.01 \\
X6 & 109.51502 & 37.866279  & $1.20\pm0.16$ & $11.2\pm1.4$ & $16.6\pm3.5$ & \emph{NE1} & 0.04 \\
X7 & 109.30196 & 37.564191 & $2.14\pm1.17$ & $6.7\pm1.2$ & $6.3\pm3.3$ & \emph{SW2} & 0.02\\
X8$^{\ast}$ & 109.54821 & 37.61343 & -- & -- & -- & -- & -- \\
X9$^{\ast\ast}$ & 109.17231 & 37.667292 & -- & -- & -- & -- & -- \\
X10$^{\ast\ast\ast}$ & 109.24965 & 37.687168 & -- & -- & -- & -- & -- \\
\hline
\hline
\end{tabular}
\label{tab_substrx}
\end{center}
\end{table*}

\subsection{X-ray \& lensing properties of substructures}
\label{sec:lensingxstr}
As already noted, MACS\,J0717's core has been extensively studied in previous work \citep{ma08,mroczkowski12,adam17a,adam17b,vanweeren17}, and we thus refer the reader to these papers for a detailed analysis of the ongoing central merger. Here we focus on the distribution of substructures in the surroundings of MACS\,J0717.

Among the 9 substructures detected in the lensing map and listed in Table~\ref{tab_substr}, two do not show a clear X-ray counterpart, \emph{SE3} and \emph{SE5}.
\emph{SE1} and \emph{SE2} are both detected in the X-ray, X1 and X2 respectively in Table~\ref{tab_substrx}. Those massive X-ray groups were already known and highlighted in \cite{jauzac12}. The alignment between X-ray and lensing peaks is almost perfect, leading to the conclusion that these two substructures are virialized, or falling in along the line of sight.
\emph{SE3} is close to X3, a complex extended X-ray substructure. While its X-ray peak aligns really well with \emph{SE4}, it is not clear that one of its components, North West of X3, could not be associated with \emph{SE3}. Indeed, this region of extended emission is apparently made of at least two and possibly three individual extended X-ray structures, as is shown by the cyan contours in Fig.~\ref{m0717_colour}. Moreover, this region is located at the edge of both the \emph{XMM-Newton} and \emph{Subaru} fields, thus uncertainties in the position of the substructures are large. For these reasons, the association of SE3 with X3 is likely.

\emph{NE1} is associated with X6, and both peaks are well aligned.
\emph{NE2} shows a bright X-ray counterpart, X4. While in Sect.~\ref{sec:lensingstr} we warned the reader that \emph{NE2} is located at the edge of the grid, the fact that in the X-ray a similar structure is detected makes us confident in that detection. However its lensing-mass estimate is biased by its proximity to the edge of the grid, and should thus be taken with care.
\emph{SW1} is almost aligned with X5. This structure exhibits a flat and elongated X-ray morphology, which could indicate a previous interaction with the main halo. However, we caution that several X-ray bright foreground substructures (labelled as X9 and X10 in Fig. \ref{m0717_colour}) are detected close to SW1 and may partly overlap with the X-ray emission associated with SW1/X5.
Concerning \emph{SW2} and its X-ray counterpart, X7, both are located at the edge of the lensing-grid and at the limit of the \emph{XMM-Newton} imaging, similar to \emph{NE2} and X5. While \emph{NE2} appears quite massive both in the lensing and X-ray maps, \emph{SW2} is the least massive substructure in our sample. It is therefore particularly difficult to disentangle between a grid artifact/edge of \emph{XMM-Newton} field of view and a real detection.
Concerning \emph{SE5}, as we explain in Sect.~\ref{sec:lensingstr}, it is located at the edge of the constrained region, therefore it could reasonably be a grid artifact, and it is also located at the edge of the \emph{XMM-Newton} field of view. Therefore we do not conclude on the existence of \emph{SE5}. In comparison with \emph{NE2}, which is clearly detected in both the lensing map and the X-ray map even if at the edge of the fields, \emph{SE5} is less massive.

\begin{table}
\caption{$M_{500}$ estimates for the secured substructures detected in the field of MACSJ\,0717.
$^{\ast}$ \emph{NE2} is located at the edge of the grid, therefore its mass estimate should be taken with care.
}
\begin{center}
\begin{tabular}[h!]{lcc}
\hline
\hline
\noalign{\smallskip}
$ID$ & $M_{500}~[10^{14}~\msun]$ \\
\hline
\emph{Core} & 4.03\\
\emph{SW1} & 1.90 \\
\emph{SE1} &  2.01 \\
\emph{SE2} & 2.07 \\
\emph{SE3} & 1.68 \\
\emph{SE4} & 1.16 \\
\emph{NE1} & 1.05 \\
\emph{NE2}$^{\ast}$ & 2.02 \\
\noalign{\smallskip}
\hline
\hline
\end{tabular}
\label{tab:nfw}
\end{center}
\end{table}

In Table~\ref{tab_substrx} we give the coordinates, the temperature, $kT$, the X-ray luminosity and the gas mass within an aperture of 250\,kpc, $L_{X,250}$ and $M_\mathrm{gas, 250}$ respectively, and for the X-ray remnant cores that have a correspondence with the lensing detections, their lensing ID, $S_\mathrm{lensing}$, as well as their gas fraction, $f_\mathrm{gas,250}$.
We note that X8, X9 and X10 do not have any lensing counterparts. To identify these substructures, we used the NED catalogue and found a corresponding object for each of them. X8 is associated with a bright spiral galaxy, 2MASSX\,J07180932+3737031, which is a GALEX source for which we could not get any redshift.
X9 is a well-known submillimetre galaxy (SMG) at $z=0.06907$, 2MASSX\,J 07164427+3739556. Finally X10 does not have any match in the NED catalogue, however we suppose it is a foreground object as there is a bright galaxy at its position. Its proximity to X9 can lead to the assumption that it can be another foreground structure at a similar redshift as 2MASSX\,J07164427+3739556.

The gas fraction within a radius of R$<250$\,kpc for all substructures with a lensing counterpart varies between 1\% and 4\%. These relatively low gas fractions can be explained by two effects. First, each of these substructures are relatively low-mass/low-temperature groups within which we do not expect the total gas fraction to exceed 10\% (see Fig.~20 in \citealt{vikhlinin06} and Fig.~4 in \citealt{eckert16b}). Second, the gas and lensing masses are measured in an aperture smaller than the virial radius of the structures, meaning that we could be missing some of the gas content and therefore we tend to underestimate the total gas fraction.

As a consistency check, we also look at the mass-temperature relation of these groups,
and compare it with the \cite{lieu16} $M$--$T$ relation, expressed as:
\begin{equation}
\log \frac{E(z)M_{500,WL}}{h_{70}^{-1}\msun} = a + b \log kT
\end{equation}
with $a=13.57$ and $b=1.67$, parameters derived from the XXL+COSMOS+CCCP sample.
The $M_{500,{\mathrm{WL}}}$ masses are estimated by fitting a NFW profile \citep{NFW97} to the integrated mass profiles we obtain for each of the substructures (see Table~\ref{tab:nfw}). Such an estimate should be considered as an upper limit, as it will tend to overestimate the mass while converting 2D projected masses into 3D masses. 
Moreover, due to the fact that substructures cannot be isolated from each other, the mass of one may contribute to the integrated mass profile of another one. 
However, while our statistics is limited, we compare our results with the $M$--$T$ relation measured by \cite{lieu16}. One of the group falls right on the \cite{lieu16} relation, and the other lie above the relation by up to a factor of 2. This suggests that the $M_{500}$ are overestimated.

\begin{table}
\caption{\label{tab:infallMXXL-dist} Evolution of radial distances in MXXL of {\sc subfind}-subhaloes for the snapshot closest to MACS\,J0717's redshift ($z \approx 0.56$), that closest to Abell\,2744's redshift ($z \approx 0.28$) and that of the closest particle data output ($z \approx 0.24$). All distances are given as radial 2D-distances from the position of the most massive subhalo in each snapshot in Mpc. 
The subhalos denoted as SH1 are the central halos.
}
 \centering
   \begin{tabular}{cccc}
 \hline\hline
    {\it ID} & $D_{z=0.56}$ & $D_{z=0.28}$ & $D_{z=0.24}$ \\
    \hline
Cluster 1 - SH 1 & 0.00 & 0.00 & 0.00 \\
Cluster 1 - SH 2 & 3.88 & 0.86 & 0.51 \\
Cluster 1 - SH 3 & 1.61 & 0.53 & 0.53 \\
Cluster 1 - SH 4 & 3.23 & 0.64 & 1.02 \\
Cluster 1 - SH 5 & 2.68 & 1.44 & 0.79 \\
Cluster 1 - SH 6 & 3.80 & 0.92 & 0.32 \\
Cluster 1 - SH 7 & 0.81 & 0.93 & 0.60 \\
Cluster 1 - SH 8 & 3.02 & 1.50 & 0.88  \\
\hline
Cluster 2 - SH 1 & 0.00 & 0.00 & 0.00 \\
Cluster 2 - SH 2 & 0.85 & 0.86 & 0.72  \\
Cluster 2 - SH 3 & 4.30 & 1.36 & 1.02  \\
Cluster 2 - SH 4 & 1.17 & 0.46 & 0.95 \\
Cluster 2 - SH 5 & 0.89 & 0.86 & 0.60 \\
Cluster 2 - SH 6 & 3.15 & 0.94 & 0.63 \\
Cluster 2 - SH 7 & 0.32 & 0.49 & 0.45 \\
    \hline\hline
  \end{tabular}
  
\end{table}%

\section{Comparison of simulations and observations}
\label{sec:compth}
Our goal is to observationally probe cluster evolution.
MACS\,J0717 is a rare object due to its mass and dynamical state at $z=0.54$. It is with such objects that we can test the limits of the cosmological paradigm.
In two previous papers \citep[][]{jauzac16b,schwinn17}, we looked at a similar cluster, Abell\,2744, at a lower redshift, $z=0.31$. Abell\,2744 has a similarly complex substructure distribution, including 7 substructures within $\sim$2 Mpc of the cluster centre (plus one background substructure identified spectroscopically, a superposition along the line of sight). Additionally 3 large-scale filaments extending out to $\sim$7\,Mpc that were detected by \cite{eckert15}. Our present analysis also finds 7 substructures around MACS\,J0717 (discarding the 2 being at the edge of the mass map and \emph{XMM-Newton} field of view), but these are farther from the cluster centre: only one (SE1) is within 2\,Mpc of the core, and the rest extend to $\sim$5\,Mpc.

\begin{table}
\caption{\label{tab:infallMXXL-mass} Evolution of masses of {\sc subfind}-subhaloes in MXXL for the snapshot closest to MACS\,J0717's redshift ($z \approx 0.56$), that closest to Abell\,2744's redshift ($z \approx 0.28$) and that of the particle data ($z \approx 0.24$). 
The mass is given as the {\sc subfind}-mass, in units of $10^{14}~\msun$.
The subhalos denoted as SH1 are the central halos.
}
 \centering
   \begin{tabular}{cccc}
 \hline\hline
    {\it ID} & $M_{z=0.56}$ & $M_{z=0.28}$ & $M_{z=0.24}$ \\
    \hline
Cluster 1 - SH 1 & 12.67 & 30.66 & 32.70\\
Cluster 1 - SH 2 & 1.62 & 0.36 & 0.17 \\
Cluster 1 - SH 3 & 0.41 & 0.10 & 0.17 \\
Cluster 1 - SH 4 & 1.24 & 0.25 & 0.22 \\
Cluster 1 - SH 5 & 0.92 & 0.36 & 0.30 \\
Cluster 1 - SH 6 & 0.19 & 0.07 & 0.02 \\
Cluster 1 - SH 7 & 0.86 & 0.19 & 0.10 \\
Cluster 1 - SH 8 & 0.89 & 0.38 & 0.23  \\
\hline
Cluster 2 - SH 1 & 14.60 & 27.30 & 29.57 \\
Cluster 2 - SH 2 & 0.10 & 0.04 & 0.03  \\
Cluster 2 - SH 3 & 3.07 & 1.46 & 0.61  \\
Cluster 2 - SH 4 & 1.22 & 0.34 & 0.40 \\
Cluster 2 - SH 5 & 0.12 & 0.06 & 0.05 \\
Cluster 2 - SH 6 & 0.01 & 0.002 & 0.01 \\
Cluster 2 - SH 7 & 0.05 & 0.04 & 0.03 \\
    \hline\hline
  \end{tabular}
  
\end{table}%

\begin{figure}
\hspace*{-5mm}\includegraphics[width=0.5\textwidth]{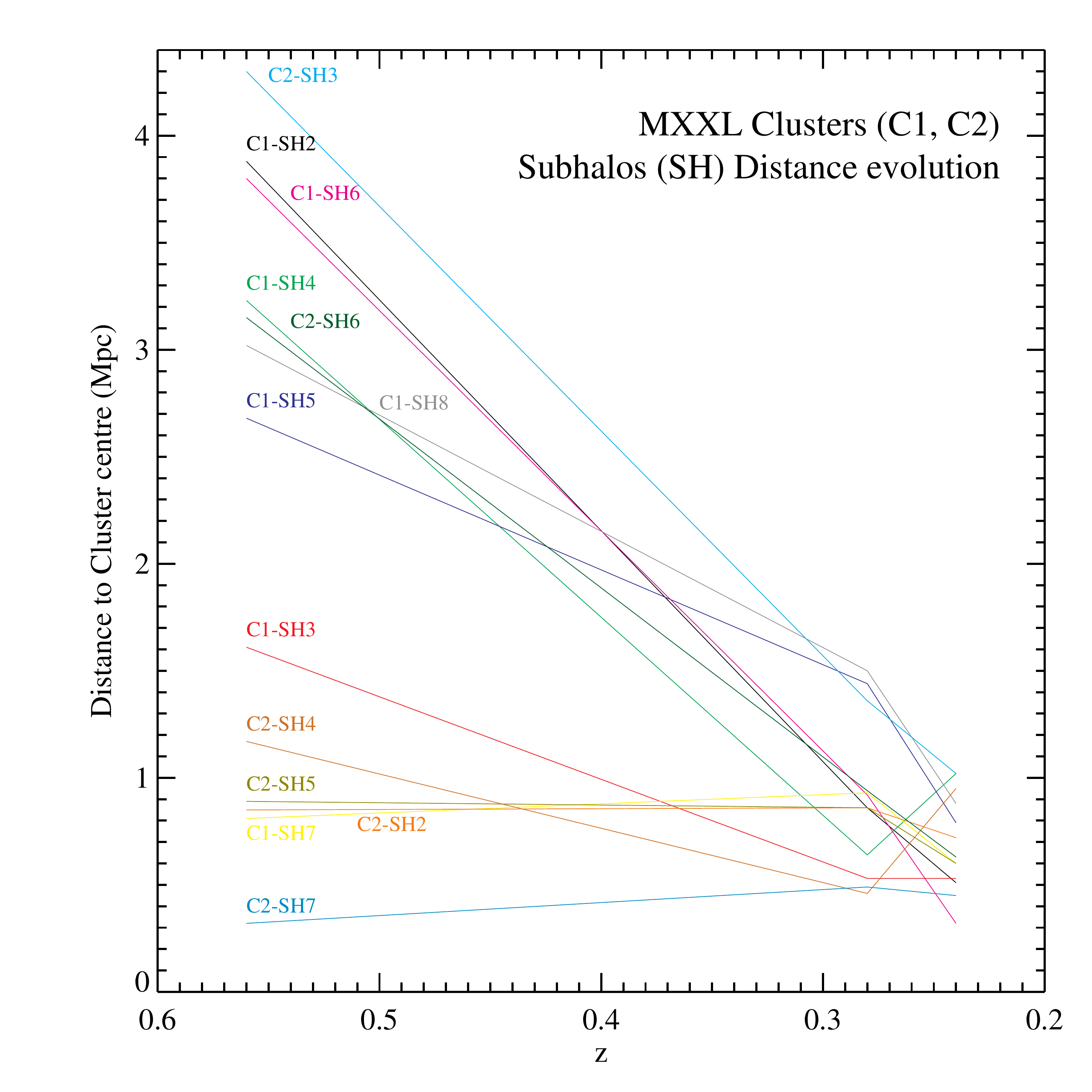}
\caption{
Projected distances to the main halo center as a function of redshift for the two MXXL clusters' subhalos as listed in Table~\ref{tab:infallMXXL-dist}. 
}
\label{distev_mxxl}
\end{figure}

Given the redshift, mass and distribution of its substructure, we hypothesize that MACS\,J0717 is the progenitor of a structure that will look very similar to Abell\,2744 by redshift $z=0.31$. To test this hypothesis, we compare our observations of these two clusters with clusters in the numerical simulations MXXL \citep[][]{angulo12} and Hydrangea/C-EAGLE \citep[][]{bahe17,barnes17}.
We shall first check whether clusters as rich in substructure as MACS\,J0717 and Abell\,2744 even exist in a $\Lambda$CDM model.
Then we shall consider the mass growth rate and substructure infall rate of similar simulated clusters, in a way that is impossible in real systems that can be observed only in a snapshot at a single redshift.

\subsection{Identification of simulated analogues}
\label{sec:simcomp}
\subsubsection{Comparison with MXXL}
We use the two halos with similar properties to Abell\,2744 presented in Sect.~\ref{sec:mxxl} to investigate the infall of substructures into
halos. 
The reason for looking for Abell\,2744 analogues rather than MACS\,J0717's is simply motivated by the lack of particle data output around $z=0.54$ with MXXL. We only have particle data at $z=0.24$, thus closer to Abell\,2744's redshift. We therefore look for Abell\,2744-like clusters (substructures close to the cluster's main halo) and trace them back in time using subhalo catalogues up to a redshift closer to MACS\,J0717's.
\emph{Cluster 1} has a
mass of $M(R < 1.3\,{\rm Mpc}) = 2.6 \times 10^{15}~\msun$ at $z=0.24$ and the
second cluster (\emph{Cluster 2}) has a mass of $M(R < 1.3\,{\rm Mpc})
= 2.5 \times 10^{15}~\msun$, both within the
3$\sigma$-range of Abell\,2744's mass. For both of these halos, we
create mass maps at $z=0.24$ by projecting all particles over a distance of $30\,
h^{-1}\, \text{cMpc}$ onto a $5 \times 5~ h^{-2}{\rm cMpc}^2$
map. Substructures within these halos are then identified as
over-densities within these maps and we check wether their mass within
an aperture of 150\,kpc lies within the $3\sigma$ interval of the
masses obtained for Abell\,2744 substructures.

Nonetheless, as mentioned in Sect.~\ref{sec:mxxl}, MXXL particle data are only available at a very small number of redshifts. If we want to analyse the evolution of the identified
subhalos up to $z=0.55$ to compare with MACS\,J0717, we are dependent
on the \textsc{subfind} datasets of all other snapshots for which the
particle data are not available. We thus identify
the \textsc{subfind}-subhalos closest to the position of each subhalo
identified in our projected mass map. We then use the merger trees
available for each \textsc{subfind}-subhalo in MXXL to trace the
evolution of each subhalo back in time, at $z=0.24$ (particle data), $z=0.28$ and finally $z=0.56$.

Table~\ref{tab:infallMXXL-dist} lists the change in radial distances
and Table~\ref{tab:infallMXXL-mass} lists \textsc{subfind}-masses of
the Abell\,2744-like substructures in both MXXL clusters. We list
distances and masses at $z = 0.56$, corresponding to the snapshot
closest to MACS\,J0717's redshift, at $z = 0.28$, the snapshot closest
to Abell 2744's redshift, and then $z = 0.24$, the snapshot where
particle data are available. For each subhalo, the radial 2D-distance
projected along the line of sight and the \textsc{subfind}-masses are
given. The analysis of these two MXXL clusters shows that subhalos
move by a distance of 2-3 Mpc between the redshifts of MACS\,J0717 and
Abell\,2744. While the subhalos already close to the virial radius do not move much, the rest of the substructures get closer to the main halo
centre by 2-3 Mpc. Figure~\ref{distev_mxxl} shows the infalling distance of subhalos as a function of redshift for \emph{Cluster 1} and \emph{Cluster 2} .

The substructures that fall in from the furthest distances correspond
mostly to the most massive substructures at redshift $z = 0.56$.
During their infall their \textsc{subfind}-masses decrease quite
dramatically, in three cases by over 70\%.  However, it is important
to be very careful when comparing \textsc{subfind}-masses to aperture
masses from gravitational lensing analysis. One reason for this
discrepancy is that \textsc{subfind} only assigns particles to a
subhalo that are gravitationally bound to it. While this is a
reasonable thing to do from a theoretical point of view, the
substructures identified by this method are not easily comparable to
those identified in gravitational lensing mass maps. In the latter,
tidally stripped material and also the background halo contribute
significantly to the subhalo masses \citep{mao17}. The degree of tidal stripping plays an important role here, which can
be seen by the fact that the mass of three subhalos drops by over
70\% (see Table~\ref{tab:infallMXXL-mass}). For a direct comparison of masses of simulated subhalos to
masses of observed subhalos, it is much more reliable to obtain the
subhalo 2D-projected masses from the simulation particle data if
available.

Finally we looked at the mass gain of the main halos of \emph{Cluster 1} and \emph{Cluster 2} considering the evolution of $M_{200}$ between $z=0.56$ and $z=0.28$.
We measure a mass growth of $M_{C1,z=0.28}/M_{C1,z=0.56}=3.7$ and $M_{C2,z=0.28}/M_{C2,z=0.56}=1.9$. If we consider the mass growth between $z=0.56$ and $z=0.00$, we obtain $M_{C1,z=0.00}/M_{C1,z=0.56}=4.0$ and $M_{C2,z=0.00}/M_{C2,z=0.56}=2.5$.

\begin{figure}
\hspace*{-3mm}\includegraphics[width=0.5\textwidth]{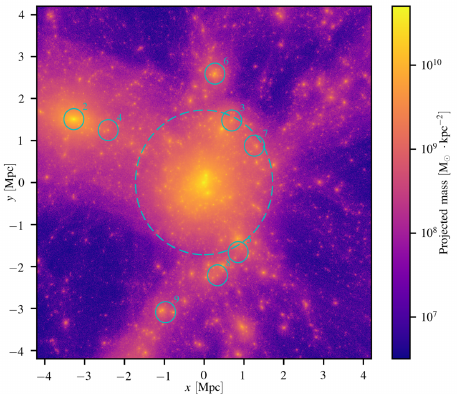}
\caption{ Projected mass map of halo CE-29 of the \ceagle simulation
  suite at $z=0.47$ centered on the cluster's minimum of potential. At
  this time the halo has a mass $M_{200}=9.15\times10^{14}~\msun$ and
  a spherical overdensity radius $R_{200}=1.72~\rm{Mpc}$ (dashed
  circle).  The small cyan circles indicate the eight subhalos that
  have a projected mass in a circular aperture of radius
  $250~\rm{kpc}$ larger than our threshold for detection (see text for
  details). The overall distribution of sub-structures is in
  qualitative agreement with what is observed in the MACSJ\,0717
  lensing mass map.}
\label{ceagle_massmap}
\end{figure}
\begin{table*}
\caption{\label{tab:infallceagle-mass} Projected mass
  ($M_{250,\rm{2D}}$) within $250~\rm{kpc}$, hot gas mass ($M_{gas,250}$) within $250~\rm{kpc}$, projected distance
  ($D_{\rm{2D}}$) to the cluster centre, velocity towards the
  centre of the halo ($v_{\rm centre}$) and the fraction of gas ($f_{gas,250}$) for the eight objects selected
  via our mock weak-lensing analysis technique in the C-EAGLE cluster CE-29 at $z=0.47$. 
  The structure designated as SH1 is the main halo.} \centering
   \begin{tabular}{cccccc|}
 \hline\hline
     \it{ID} & $M_{250,\rm{2D}}$ & $M_{gas,250}$ & $D_{\rm{2D}}$ & $v_{\rm centre}$ & $f_{gas,250}$ \\
      & $[10^{13}\msun]$ & $[10^{11}\msun]$ & $[\rm{Mpc}]$ & $[\rm{km}/\rm{s}]$ & \\
    \hline
CE29 - SH 1 & 20.1 & 59.8 & 0 & 0 & 0.03 \\
CE29 - SH 2 & 9.16 & 48.9 & 3.60 & 440 & 0.05 \\
CE29 - SH 3 & 1.64 & 0.9 &1.63 & -1100 & 0.01 \\
CE29 - SH 4 & 1.09 & 2.7 & 2.71 & 290 & 0.03 \\
CE29 - SH 5 & 1.02 & 0.4 & 1.87 & 300 & 0.004 \\
CE29 - SH 6 & 1.47 & 6.4 & 2.61 & 1070 & 0.04 \\
CE29 - SH 7 & 1.04 & 0.8 & 1.54 & 1290 & 0.01 \\
CE29 - SH 8 & 1.05 & 0.2 & 2.24 & 1280 & 0.002 \\
CE29 - SH 9 & 1.01 & 0.004 & 3.24 & 950 & $4.0\times10^{-5}$ \\
    \hline\hline
  \end{tabular}
  
\end{table*}%

\begin{figure}
\hspace*{-5mm}\includegraphics[width=0.5\textwidth]{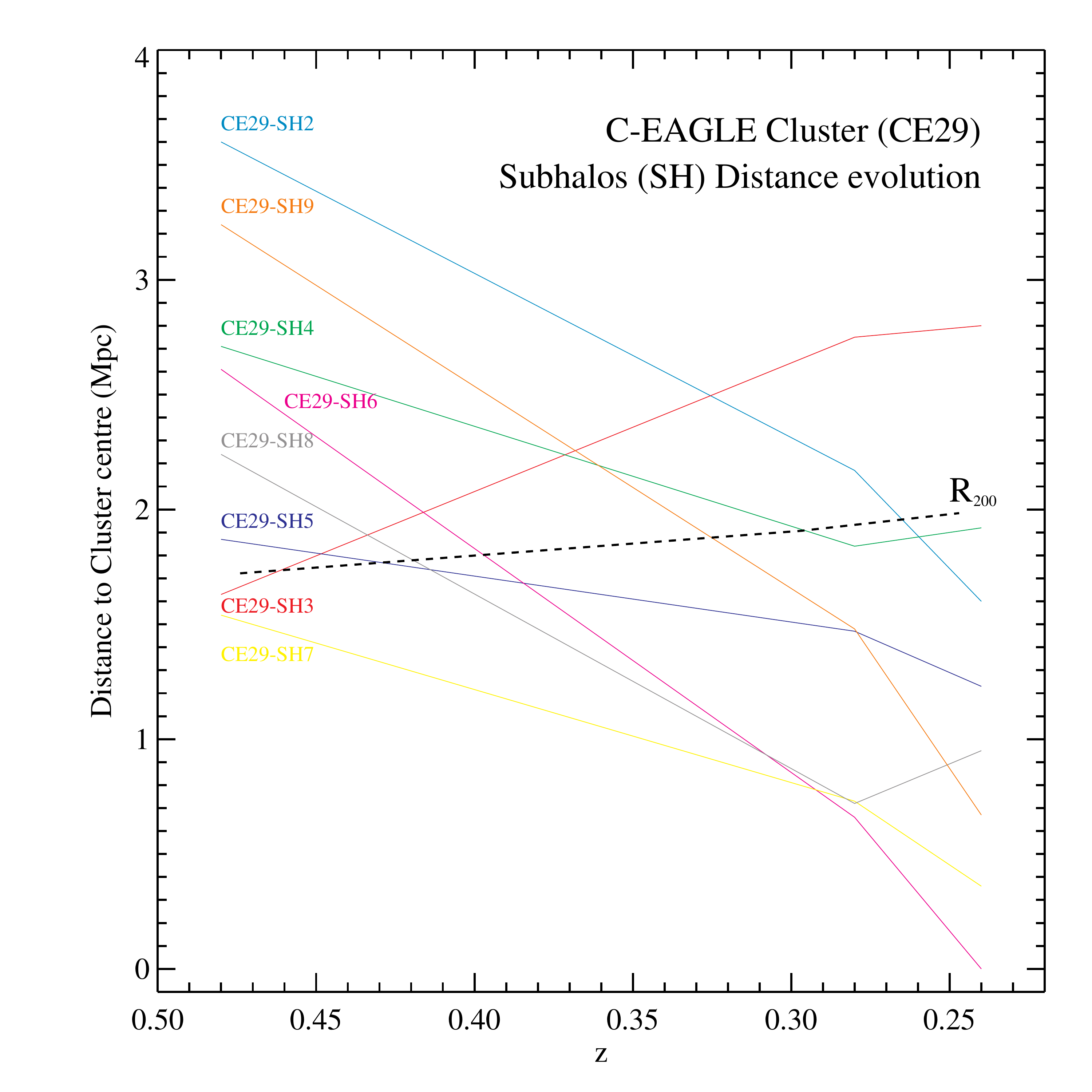}
\caption{
Projected distances to the main halo center as a function of redshift for the C-EAGLE cluster' subhalos as listed in Table~\ref{tab:infallceagle-dist}. the black dashed line represents the evolution of $R_{200}$. As one can see, all substructures except for CE29-SH2 are within the virial region by $z=0.24$.
}
\label{distev_ceagle}
\end{figure}

\subsubsection{Comparison with Hydrangea/C-EAGLE}
For the purpose of our comparison, we use the most massive halo (CE-29) present in the simulation. At $z=0$
this cluster has a radius $R_{200} = 2.8~\rm{Mpc}$, a mass
$M_{200}=2.4\times10^{15}~\msun$, a soft X-ray luminosity of
$L^{0.5-2.0{\rm keV}}_{500}=8.8\times10^{44}~{\rm erg}{\rm s^{-1}}$,
a spectroscopic temperature $k_{\rm B}T_{500}=7.7~\rm{keV}$ and is
the host of $826$ galaxies with a stellar mass $>10^9~\msun$ \citep{bahe17,barnes17}.

We analyse the snapshot of the simulation closest to the redshift of MACS\,J0717 at $z=0.47$. We extract a cube of side-length $10~\rm{Mpc}$
centered around the minimum of the potential of the halo. To construct a
mass map, we project all the particles along the z-axis of the
simulation volume and bin them using a regular grid with cells of
side-length $2~\rm{kpc}$. The result of this procedure is shown in
Fig.~\ref{ceagle_massmap} with the dashed circle corresponding to the
projected spherical over-density radius
$R_{200}=1.72~\rm{Mpc}$. Compared to actual weak-lensing data, this
projected mass map is an idealised mass map where any foreground and
background structures perturbing the signal have been removed.

\begin{table}
\caption{\label{tab:infallceagle-massev} Masses of the C-EAGLE cluster CE-29 subhaloes at $z=0.48$, $z=0.29$ and $z=0.24$. All masses are \textsc{subfind} masses and are given in $\msun$. The structure designed as SH1 is the main halo.
}
 \centering
   \begin{tabular}{lcccc}
 \hline\hline
    {\it ID} & $M_{z=0.48}$ & $M_{z=0.29}$ & $M_{z=0.24}$\\
    \hline
CE29 - SH 1 & 9.15$\times10^{14}$ & 9.76$\times10^{14}$ & 10.70$\times10^{14}$ \\
CE29 - SH 2 & 2.07$\times10^{14}$ & 1.89$\times10^{14}$ & 1.57$\times10^{14}$\\
CE29 - SH 3 & 1.68$\times10^{12}$ & 1.37$\times10^{12}$ & 1.32$\times10^{12}$\\
CE29 - SH 4 & 6.46$\times10^{11}$ & 5.43$\times10^{11}$ & 4.74$\times10^{11}$\\
CE29 - SH 5 & 5.55$\times10^{11}$ & 4.56$\times10^{11}$ & 4.55$\times10^{11}$ \\
CE29 - SH 6 & 1.48$\times10^{13}$ & 1.46$\times10^{9}$ & \emph{merged}\\
CE29 - SH 7 & 4.23$\times10^{10}$ & 2.82$\times10^{10}$ & 1.63$\times10^{10}$\\
CE29 - SH 8 & 3.13$\times10^{11}$ & 2.37$\times10^{11}$ & 2.76$\times10^{11}$\\
CE29 - SH 9 & 3.13$\times10^{10}$ & 3.22$\times10^{10}$ & 1.19$\times10^{10}$\\
    \hline\hline
  \end{tabular}
\end{table}

\begin{table}
\caption{\label{tab:infallceagle-dist} Radial distances of the C-EAGLE cluster CE-29 subhaloes at $z=0.47$, $z=0.30$ and $z=0.25$. All distances are given as radial 2D-distances from the position of the main halo (SH1) in each snapshot in Mpc.
}
 \centering
   \begin{tabular}{lcccc}
 \hline\hline
    {\it ID} & $D_{z=0.47}$ & $D_{z=0.30}$ & $D_{z=0.25}$\\
    \hline
CE29 - SH 1 & 0.00 & 0.00 & 0.00 \\
CE29 - SH 2 & 3.60 & 2.17 & 1.60\\
CE29 - SH 3 & 1.63 & 2.75 & 2.80\\
CE29 - SH 4 & 2.71 & 1.84 & 1.92\\
CE29 - SH 5 & 1.87 & 1.47 & 1.23 \\
CE29 - SH 6 & 2.61 & 0.66 & 0.00\\
CE29 - SH 7 & 1.54 & 0.73 & 0.36\\
CE29 - SH 8 & 2.24 & 0.72 & 0.95\\
CE29 - SH 9 & 3.24 & 1.48 & 0.67\\
    \hline\hline
  \end{tabular}
\end{table}

We then construct a catalogue of weak-lensing detected
objects. We start by selecting all the haloes and sub-haloes
identified by the \subfind algorithm with a mass above $10^{10}~\msun$
and compute the total projected mass in a $250~\rm{kpc}$ circular
aperture around their centre of potential. In a second step we discard
all such sub-structures with a projected mass under
$10^{13}~\msun$. This is slightly lower than the smallest object
detected around MACS\,J0717 but is designed to allow for a
systematical overestimation of the masses in the weak-lensing data. As
pointed out by \cite{schwinn17} and more quantitatively
by \cite{mao17}, \subfind masses two orders of magnitude lower than a
given aperture mass can be boosted by projection effects to reach the mass
threshold. As we are aiming for a projected aperture mass of
$10^{13}\msun$, our first selection of sub-haloes with
a \textsc{subfind} mass above $10^{10}$ is justified. This procedure,
however, does not guarantee that the structures analysed in this way
would be detectable. An additional step to identify detectable 
overdensities is required. We hence iterate over all the substructures
from the most massive to the least massive and eliminate the ones that
overlap with a more massive object or that are within $1~\rm{Mpc}$ of
the centre of the halo. This step is necessary as there is no way
observationally to distinguish structures that overlap in
projection. At the end of this procedure, we are left with 8
sub-haloes shown as small cyan circles on Fig.~\ref{ceagle_massmap}
with their masse, projected distance to the centre and velocity
towards the centre of the cluster given in the first three columns of
Table~\ref{tab:infallceagle-mass}. The mass and distance range is similar to what we observe in MACS\,J0717. We also analysed the companion
simulation without baryon physics and found projected masses in
excellent agreement. This demonstrates that baryonic processes have
little effects on these weak lensing mass measurements.

From the total masses of the halo ($M_{200}$), we can estimate a mass growth rate  between $z=0.47$ and $z=0.29$ of $M_{z=0.47}/M_{z=0.29}=1.20$. Moreover, between $z=0.47$
and $z=0.00$ we measure a mass growth rate $M_{z=0.47}/M_{z=0.00}=2.62$. We note that
CE-29 is undergoing a major merger at $z=0.24$, which is responsible for such a high mass growth. 

In Table~\ref{tab:infallceagle-massev} and Table~\ref{tab:infallceagle-dist}
 we show the evolution of the
projected distance of the subhalos and their \textsc{subfind} masses
between $z=0.47$ and $z=0.24$. This is the closest we can get with
available particle data to the respective redshifts of MACS\,J0717 and
Abell\,2744.  As one can see in Fig.~\ref{distev_ceagle}, they all
have entered the virial radius region, except for CE29 - SH3 which is
a back-splash sub-halo that will return towards the centre of the
cluster at a later time. The average projected distance travelled to the main
halo between $z=0.47$ and $z=0.24$ is in the range of 1-2\,Mpc as can
be seen in Fig.~\ref{distev_ceagle}.

As discussed above, the \textsc{subfind} masses cannot be related in a
straightforward way to the projected mass within $250~\rm{kpc}$ that
can be observationally measured. \cite{mao17} showed that sub-halos
embedded in a large halo can see their ratio of projected mass
over \subfind mass reach values as large as $10^3$. Similar boosts can
also be seen in the substructures detected outside the virial radius
but part of the larger over-density around the cluster. This can be
seen by comparing the masses reported in
Table~\ref{tab:infallceagle-mass} and
Table~\ref{tab:infallceagle-massev}.

\subsection{Properties of a high redshift super-cluster}
\label{sec:formclus}
\subsubsection{Infall of Substructures}
MACSJ\,0717 substructures seem to be distributed along three preferred directions: South-East, North-East and South-West, with neither lensing nor X-ray substructures detected in the North-West region as can be seen in Fig.~\ref{m0717_colour}. In the context of hierarchical structure formation scenarios, halos are expected to be located at the intersection of three large-scale filaments \citep{bond96}, as was observed around Abell\,2744 \citep{eckert15}. 
Nevertheless, all the dark matter substructures along with their X-ray counterparts, apart from the \emph{Core} and \emph{SE1}, appear to be located outside the virial radius of MACS\,J0717 (yellow circle on Fig.~\ref{m0717_colour}).
However, \cite{lau15} show that up to $\sim$3$\times$R$_{200}$ ($\sim$7\,Mpc), these substructures are already decoupled from the Hubble flow and therefore infalling into the cluster. Therefore, looking at both the dark and luminous mass distribution of MACS\,J0717, it is reasonable to assume that these detected substructures are falling into the cluster's main halo, and are doing so along three preferred directions, one of them at the location of which a dark matter large-scale filament was detected by \cite{jauzac12}.
However, only higher quality X-ray and weak-lensing data will enable us to confirm this assumption. 

\begin{figure*}
\hspace*{-5mm}\includegraphics[width=\textwidth]{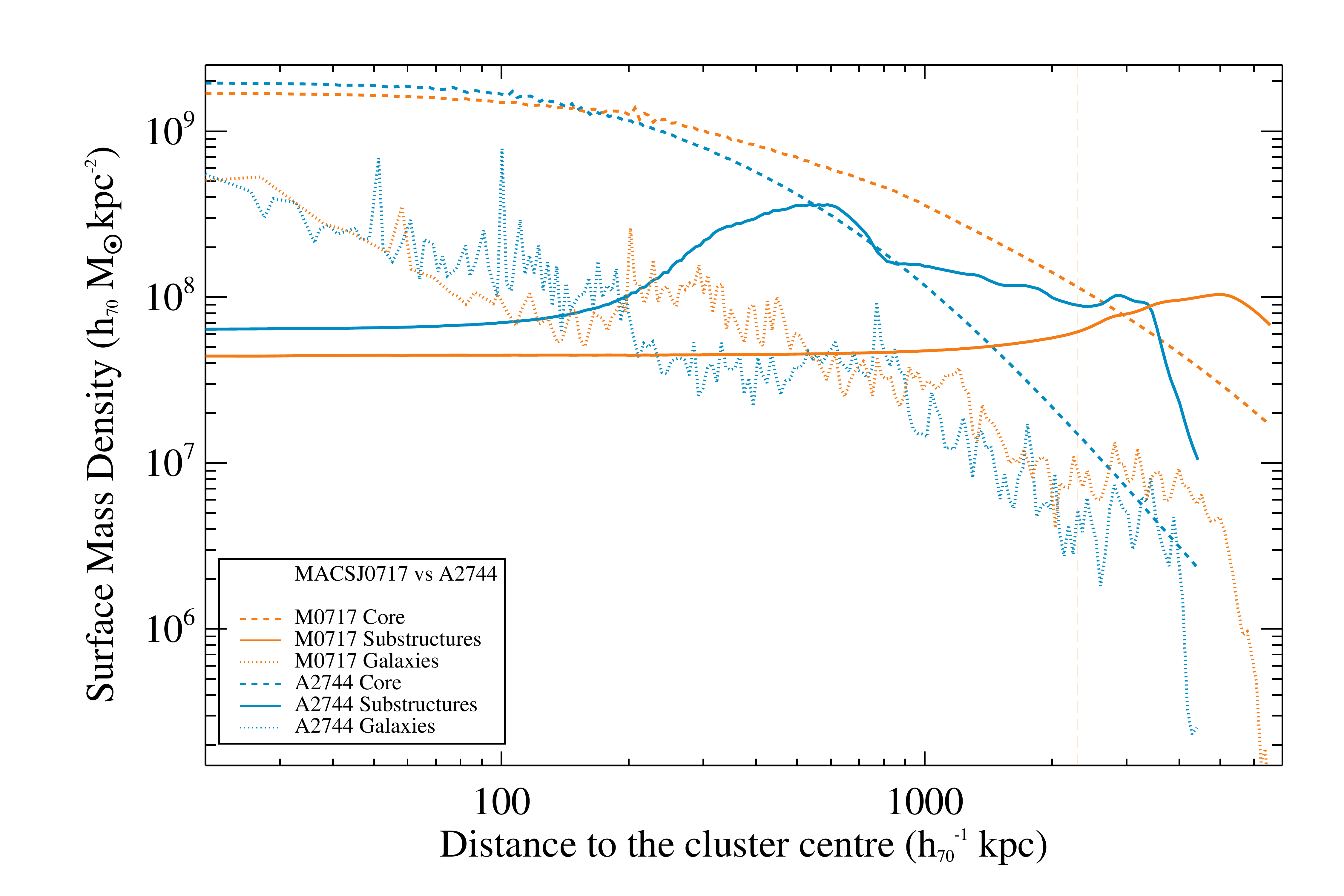}
\caption{
Density profile of the different components in MACS\,J0717 (orange) and Abell\,2744 (cyan) as a function of the distance to the cluster center: the core component (dashed line), the galaxy component (dotted lines) and the substructure component (plain lines). 
We differentiate between galaxies and (massive) substructures.
The (long) dashed vertical lines highlight R$_{200}$ for both clusters, MACS\,J0717 and Abell 2744 in orange and cyan respectively.
}
\label{densprof_compo}
\end{figure*}

To start our investigation, we first estimate the distance that substructures like the ones observed here, would travel within the redshift interval between MACS\,J0717 and Abell\,2744.
For this we calculate the interval in look-back time, $\Delta t_{lb}$, for both cluster redshifts using the following expression:
\begin{equation}
\label{eq:lookback}
t_{lb}(z) = t_{H} \int_{0}^{z} \frac{dz'}{(1+z')} E(z')\ ,
\end{equation}
where,
\begin{equation}
E(z) = \sqrt{(\Omega_{m}(1+z)^3 + \Omega_\Lambda}\ .
\end{equation}
Applying equation~\ref{eq:lookback} to the redshifts of Abell\,2744 and MACS\,J0717 we obtain $t_{lb}(z=0.31) = 3.4\times 10^{9}$\,yr, and $t_{lb}(z=0.54) = 5.3\times 10^{9}$\,yr. That gives us a look back time $\Delta t_{lb} =2.1 \times 10^{9}$\,yr.
If we make the hypothesis that the observed substructures in MACS\,J0717 are typical groups which have an infall velocity of $\sim1000$\,km.s$^{-1}$ \citep{lau15}, thus we estimate that between $z=0.54$ and $z=0.31$, the typical infall distance of substructures should be $\sim$2\,Mpc. This value is a lower limit estimate due to both projection effects and the fact that once substructures have entered the virial radius region, their infall velocity will increase as they are getting closer to the centre of the main halo. 

Figure~\ref{densprof_compo} shows the density profiles of both MACS\,J0717 (orange) and Abell\,2744 (cyan) as a function of the distance to the cluster centre for each of their components: the core of the cluster (dashed line), the substructure contribution (plain line) and the contribution from galaxies (dotted lines). We also highlight their respective $R_{200}$ as vertical lines ($R_{200,0717}=2.3$\,Mpc; $R_{200,2744}=2.1$\,Mpc).
From the core profiles, one can clearly see the different evolution stages between the two clusters: MACS\,J0717 has a slightly less dense but more extended core than Abell\,2744, a typical signature of the change of the mass-concentration relation with redshift and mass. With time, substructures will infall and merge, the core density will increase and become more `peaky' while leaving the outskirts of the cluster slightly under-dense as can be seen in Abell\,2744.
While substructures often refer to galaxies, we here refer to group to cluster-scale halos ($M>3.5\times10^{13}~\msun$). The substructure profiles in Fig.~\ref{densprof_compo} clearly show the different evolutionary stages of the two clusters: the density of the substructure distribution peaks around $\sim$5-6\,Mpc in MACSJ\,0717, and peaks at $\sim$1\,Mpc followed by a plateau up to 4.5\,Mpc (limit of the field of view) in Abell\,2744. This change of slope in Abell\,2744 substructure's density profile is due to the presence of the large-scale filaments detected by \cite{eckert15}. 
As a matter of consistency, we show all the clusters' components in Fig.~\ref{densprof_compo}. While the evolution stage of the object plays a key role in the shape of the density profiles of both the core and the substructure profiles, one can see that the galaxy density profiles for both clusters are similar and follow a similar slope.
Applying our above infall distance estimate between $z=0.54$ and $z=0.31$ to MACS\,J0717 would mean that all the substructures would reach the virial radius by $z=0.31$.

However, this analytic approach is limited. That is why we turn to numerical simulations such as MXXL \citep{angulo12} and Hydrangea/C-EAGLE \citep{bahe17,barnes17} in order to trace these `independent' substructures (i.e. at $R>R_{200}$) between $z\sim0.55$ and $z\sim0.3$ and measure their average infall distance.
In Abell\,2744, all the substructures are detected within less than 2\,Mpc from the core (excluding the filamentary structures outside $R_{200}$). Therefore all substructures are assumed to be virialized within the main halo, which is not the case for MACS\,J0717.
As explained at the beginning of this Section, one motivation behind this analysis is to see if the assumption that MACS\,J0717 could be a progenitor of an Abell\,2744-like cluster is realistic in terms of substructure infall and thus distribution. According to the calculation made earlier, it appears to be sensible (considering $\sim$2\,Mpc as a lower limit of infall distance between the redshift of the two clusters) to postulate that the actual substructures visible in MACS\,J0717 outskirts would have reached $R_{200}$ by redshift $z=0.31$, and is in good agreement with both MXXL and Hydrangea/C-EAGLE.

\subsubsection{Mass growth rate}

We now estimate the mass growth of MACS\,J0717.
In order to estimate the growth due to substructure infall, we fit a NFW profile to all the secured substructures (see Sect.~\ref{sec:lensingxstr} and Table~\ref{tab:nfw}), i.e. excluding \emph{SE5} and \emph{SW2} from our calculations as well as \emph{SE1} which is already in the main halo ($D_{C-SE1}=1.6$\,Mpc).
We can thus estimate a total mass of substructures $M_{500,{\rm Sub}}=$0.98$\times$10$^{15}~\msun$. Applying a similar fit to the \emph{Core} component, we obtain a mass $M_{500,Core}=$4.03$\times$10$^{15}~\msun$. 
Therefore, considering that all these substructures will be merging with the cluster main halo within $R_{200}$ at $z=0.31$, we can estimate that the mass growth of MACS\,J0717 due to massive substructures will be of a factor of 1.25 (over $\Delta t_{lb} =2.1 \times 10^{9}$yr). 

When we compare the mass growth of MACS\,J0717 due to these substructures only we can see a good agreement with the mass growth estimated from Hydrangea/C-EAGLE ($M_{z=0.48}/M_{z=0.29}=1.20$) and slightly lower than MXXL measurements ($M_{C1,z=0.28}/M_{C1,z=0.56}=3.7$ and $M_{C2,z=0.28}/M_{C2,z=0.56}=1.9$). 
Nevertheless, a few caveats have to be emphasized: (1) the MACS\,J0717 calculation relies on our conversion of aperture projected masses into $M_{500}$, and thus may lead to an overestimation of the mass growth rate, and (2) the growth rate relies on projected masses, however as shown by \cite{mao17} such masses can be overestimated by up to 2 orders of magnitude. 
Our relative agreement with numerical simulations could lead to the conclusion that the mass growth rate of these `cosmic beasts' is not dominated by the smooth accretion of surrounding material (low mass substructures, i.e. not groups nor clusters), but rather by events of massive substructure infall. 
However, the true masses of the simulated substructures are given in Table~\ref{tab:infallMXXL-mass} and Table~\ref{tab:infallceagle-massev}. We note that the difference between those and their projected estimates can reach several order of magnitudes which is in agreement with \cite{mao17}. That is suggesting that the majority of the mass growth measured for the simulated clusters is due to smooth accretion of low mass substructures rather than infall of massive substructures. 

MACS\,J0717's substructure distribution (up to $R\sim5$\,Mpc) and total mass leads us to the conclusion that we are observing a super-cluster \citep{einasto01,einasto07,chon13}, similar to what is observed in \cite{pompei16} at $z=0.43$. If by $z=0$ MACS\,J0717 has virialized, it will form an extremely massive cluster of $M_{200}\sim10^{16}\,\msun$ considering the average mass growth rate of 2.9 measured from both MXXL and C-EAGLE clusters -.
The clusters considered with MXXL and Hydrangea/C-EAGLE are the most massive objects visible in the simulations, and are all three undergoing extreme merger events between $0.2<z<0.3$, making the average growth rate relatively large. However, considering the complexity of MACS\,J0717, we could expect it to undergo such an extreme dynamical history.

\subsubsection{Gas fraction in substructures}
Finally, we took advantage of the Hydrangea/C-EAGLE simulation that includes baryons in order to compare our measured gas fractions with the ones of the CE-29 substructures. The gas
mass, $M_{gas,250}$, of each CE-29 substructures measured within an aperture
of 250\,kpc and the gas fraction, $f_{gas,250}$, are reported in
Table~\ref{tab:infallceagle-mass}. We here consider the hot gas mass
in order to compare our results with what is measured in
MACS\,J0717. 

We measure relatively low gas fractions, between 1\% and 5\%. This is in excellent
agreement with MACS\,J0717 gas fractions as reported in
Table~\ref{tab_substrx}. It confirms our initial hypothesis of
an underestimated gas fraction due to a small aperture (not extending
up to $R_{200}$), as well as the reliability of the baryonic physics in the Hydrangea/C-EAGLE suite of simulations.

\section{Conclusions}
\label{sec:conclusion}
The masses and distribution of substructures in the outskirts of massive galaxy clusters provide an observational test of the $\Lambda$CDM structure formation paradigm and at the same time present an opportunity to quantify the importance of major mergers in the buildup of these extreme objects. We have performed a combined strong-, weak-lensing, and X-ray analysis of observations from the \emph{HST}, Subaru, \emph{Chandra}, and \emph{XMM-Newton} telescopes. We detected substructures in the outskirts of one of the most massive galaxy clusters in the observable Universe, MACS\,J0717 at $z=0.54$, using the hybrid version of the \textsc{lenstool} software \citep{jullo07,jullo10,jauzac12,jauzac15a}. To interpret our findings, we have compared our observational results to our previous analysis of the massive cluster Abell 2744 at $z = 0.31$ \citep{jauzac16b} and two different cosmological simulations, MXXL \citep{angulo12} and Hydrangea/C-EAGLE \citep{bahe17,barnes17}.

\paragraph*{Observational results}
Our key observational results may be summarized as follows:
\begin{enumerate}
\item We detect 9 group-scale substructures with masses ranging between $M(R<250~kpc)=3.8-6.5\times10^{13}~\msun$ located between 1.6 and 4.9\,Mpc from the cluster centre. 

\item The X-ray analysis of the \emph{XMM-Newton} and \emph{Chandra} data reveals 10 substructures.

\item The combination with X-ray data allowed us to secure 7 of the lensing detections.

\item The X-ray data show 3 substructures not detected in the lensing mass map (X8, X9 and X10) that we identified as being possible foreground objects in the NED catalogue. 

\item We measure the fraction of gas within a radius of 250\,kpc, $f_{gas,250}$, for all substructures. $f_{gas,250}$ varies between 1\% and 4\%. 
This is well below the cosmic mean, in agreement with previous studies \citep{vikhlinin06,eckert16}.

\item We look at the $M_{500,WL}$-T relation for these groups and compare our results with \cite{lieu16}. This confirms the overestimation of our $M_{500}$.
\end{enumerate}

\paragraph*{Comparison with MXXL \ Hydrangea/C-EAGLE}
Our key results from the comparison with numerical simulations can be summarized as follows:
\begin{enumerate}

\item Clusters as rich in substructure as MACS\,J0717 and Abell 2744 are common in cosmological simulations, if the simulation data are analysed in a way compatible to the observations.

\item Projected mass maps of the two most massive clusters in the MXXL dark-matter only simulation, and the most massive cluster in the hydrodynamical Hydrangea/C-EAGLE simulation (CE-29) all reveal a similar number of massive substructures as what is observed. The substructures in the latter also have hot gas fractions that are in excellent agreement with MACS\,J0717 (Figs.~\ref{distev_mxxl}, \ref{ceagle_massmap}, and \ref{distev_ceagle}).

\item From the total halo mass of the two MXXL clusters, we can estimate a mass growth rate of a factor of 3.7 and 1.9 for \emph{Cluster 1} and \emph{Cluster 2} respectively between $z=0.56$ and $z=0.28$, and of a factor of 4.0 and 2.5 between $z=0.56$ and $z=0.00$.

\item For the first time we confronted theory and observations of gas fractions in massive galaxy clusters using state-of-the-art hydrodynamical simulations. We measured the gas fraction within an aperture of 250\,kpc for each of CE-29 substructures, between 1\% and 5\%, which is in excellent agreement with what is observed in MACS\,J0717. From the evolution of the total halo mass in the Hydrangea/C-EAGLE cluster, $M_{200}$, we estimated a mass growth of a factor of 1.2 between $z=0.48$ and $z=0.24$.
\end{enumerate}

\paragraph*{Mass growth of MACS\,J0717}
The substructure distribution in MACS\,J0717 is quite similar to what we observed in Abell\,2744: seven substructures at the cluster redshift detected in both cases, with relatively lower masses in the case of MACS\,J0717 and more distant to the cluster centre. Such behavior is expected while looking at the redshift difference between the two clusters, one being at a more advanced evolution stage than the other. Therefore we ask ourselves the question of wether MACS\,J0717-like clusters coud be the progenitors of Abell\,2744-like clusters:
\begin{enumerate}

\item A comparison between MACS\,J0717 and Abell\,2744 suggests that the substructures we have identified in the former will move in towards the cluster centre by $\sim$2--3 Mpc between $z = 0.54$ and $z = 0.31$. This agrees with both analytic expectations and the radial motion of substructures in the simulations, and suggests that MACS\,J0717 will, over time, evolve into a system similar to Abell\,2744 in terms of substructure distribution.

\item Compared to Abell\,2744, the core of MACS\,J0717 shows a more extended mass profile of its core and substructure components, while the mass profiles from their individual member galaxies agree well. We interpret this as evidence for a less evolved state of MACS\,J0717, as might be expected from its higher redshift and mass (Fig.~\ref{densprof_compo}). 

\item From the lensing masses and expected infall velocity of the MACS\,J0717 substructures, we estimate a mass growth due to the accretion of massive substructures of a factor of 1.25 between $z = 0.54$ and $z = 0.31$, i.e.~over a period of 2.1\,Gyr. This is close to the total growth of the simulated cluster CE-29 over a similar redshift interval (1.20), but lower than the growth of either of the two MXXL clusters (3.7 and 1.9, respectively). Taking into account that our lensing-derived substructure masses are likely overestimated, this implies that the growth of `cosmic beast' is dominated by the smooth accretion of surrounding material (low-mass substructures) rather than regular massive substructure infall events.
\end{enumerate}

\paragraph*{A super-cluster at ${\bf z=0.54}$}
MACS\,J0717 is a super-cluster at $z=0.54$ \citep{einasto01,einasto07,chon13,pompei16}. Extrapolating its mass growth with expectations from simulations, it has likely evolved into an extremely massive cluster of $M_{200} \approx 10^{16}\,\mathrm{M}_\odot$ by the present day. We have shown that such massive systems are commonly surrounded by a large number of group-scale substructures, in agreement with recent observations at lower redshift \citep{haines17} and cosmological simulations in a $\Lambda$CDM cosmology. Rather than constituting a challenge to our understanding of structure formation, our study has demonstrated that such objects offer a unique opportunity to directly observe the assembly of massive galaxy clusters. In the near future, we can hope to exploit this further by analyzing the substructure content of other massive clusters, and extending it to proto-clusters at higher redshift. Combined with cosmological simulations, such observations will enable a detailed probe of the assembly of `cosmic beasts', the most massive bound structures in the Universe.

\section*{Acknowledgements}
We thank the anonymous referee for their comments which helped improve our paper.

\noindent This work would not have been possible without Lydia Heck and Peter
Draper's computing expertise. We thank John Helly for his help with
merger trees, and R.\ Kawamata and J.\ M.\ Diego for sharing their 
integrated mass profiles.

\noindent 
This work was supported by the Science and Technology
Facilities Council (grant numbers ST/L00075X/1, ST/P000451/1) and used
the DiRAC Data Centric system at Durham University, operated by the
Institute for Computational Cosmology on behalf of the STFC DiRAC HPC
Facility (\url{www.dirac.ac.uk}). This equipment was funded by BIS
National E-infrastructure capital grant ST/K00042X/1, STFC capital
grant ST/H008519/1, and STFC DiRAC Operations grant ST/K003267/1 and
Durham University. DiRAC is part of the National E-Infrastructure.
The Hydrangea/C-EAGLE simulations were in part performed on the German
federal maximum performance computer ``HazelHen'' at the maximum
performance computing centre Stuttgart (HLRS), under project GCS-HYDA
/ ID 44067 financed through the large-scale project ``Hydrangea'' of
the Gauss Center for Supercomputing.  Further simulations were
performed at the Max Planck Computing and Data Facility in Garching,
Germany. 
This project has received funding from the European Union's Horizon 2020 research and innovation programme under the Marie Sklodowska-Curie grant agreement No 747645.
MJ, EJ and ML acknowledge the support of the Centre National d'\'Etudes Spatiales (CNES).
MJ, ML, and EJ acknowledge the M\'esocentre d'Aix-Marseille Universit\'e (project number: 14b030). This study also benefited from the facilities offered by CeSAM (Centre de donn\'eeS Astrophysique de Marseille (\url{http://lam.oamp.fr/cesam/}). 
DE acknowledges the funding support from the Swiss National Science Foundation (SNSF).
RM is supported by the Royal Society.
CDV acknowledges financial support from the Spanish Ministry of Economy and Competitiveness (MINECO) through grants AYA2014-58308 and RYC-2015-1807.
JPK and MJ acknowledge support from the ERC advanced grant LIDA. 
ML acknowledges the Centre National de la Recherche Scientifique (CNRS) for its support. 
MN acknowledges PRIN INAF 2014 1.05.01.94.02.
PN acknowledges support from the National Science Foundation via the grant AST-1044455, AST-1044455, and a theory grant from the Space Telescope Science Institute HST-AR-12144.01-A. 
KU acknowledges support from the Ministry of Science and Technology of Taiwan through the grant MoST 106-2628-M-001-003-MY3.

\noindent {\it Facilities:} This paper uses data from Hubble Space Telescope programmes GO-09722, GO-11560, GO-12103, GO-13498. 




\bibliographystyle{mnras}
\bibliography{reference} 



%
%


\bsp	
\label{lastpage}
\end{document}